\documentclass[12pt]{article}

\usepackage[dvips]{graphics}
\usepackage{amssymb,amsmath}
\usepackage{color}
\usepackage{graphicx}
\usepackage[square,numbers]{natbib}   
\newcommand{\beq}[1]{  \begin{equation} \label{#1} }
\newcommand{\eeq}{     \end{equation}}
\newcommand{\bal}[1]{\begin{align} \label{#1} }

\def\Re{\mathrm{Re}}
\def\Im{\mathrm{Im}}
\def\dd{\mathrm{d}}
\def\tr{\mathrm{tr}}
\def\diag{\mathrm{diag}}

\def\sign{\operatorname{sign}}
\def\ee{\mathrm{e}}
\def\div{\mathrm{div}}
\def\nd{m}
\def\z{{\mathbf z}}
\def\o{}

\def\singlespacing{\baselineskip=13pt}	

\begin{document}
\singlespacing

\title{\color{blue} 
Wave impedance matrices  for  cylindrically anisotropic  \\ radially inhomogeneous  elastic solids }
\author{A. N. Norris\footnote{Rutgers University, Department of Mechanical and
Aerospace Engineering, 98 Brett Road, Piscataway, NJ  08854-8058,
norris@rutgers.edu} $\,$ and A. L. Shuvalov\footnote{Universit\'e de Bordeaux, CNRS, UMR 5469, Laboratoire de M\'ecanique Physique, 351 Cours
de la Lib\'eration, Talence F-33405, France} }\maketitle



\begin{abstract}

  Impedance matrices are obtained for radially inhomogeneous structures
using  the Stroh-like system of six first order differential equations for the time harmonic displacement-traction 6-vector. 
Particular attention is paid to the newly identified solid-cylinder impedance matrix ${\mathbf Z} (r)$ appropriate to cylinders with material at $r=0$, and its limiting value at that point,  
the  solid-cylinder impedance
matrix ${\mathbf Z}_0$.       
We show that ${\mathbf Z}_0$ is   a fundamental material property depending only on the elastic moduli and the azimuthal order $n$,  
that  ${\mathbf Z} (r)$ is  Hermitian and ${\mathbf Z}_0$ is negative semi-definite.   Explicit solutions for ${\mathbf Z}_0$ are presented for monoclinic and higher material symmetry, and  the 
  special cases of  $n=0$ and $1$ are treated in detail. 
Two methods are proposed for finding ${\mathbf Z} (r)$, one based on the Frobenius series solution and the other using  a differential Riccati equation  with   ${\mathbf Z}_0$ as initial value. 
The radiation impedance matrix is   defined and   shown to be non-Hermitian. 
These  impedance matrices enable  concise and efficient  formulations of  dispersion equations for wave guides,   and solutions of scattering and related    wave problems in  cylinders.

 \end{abstract}

 \section*{Introduction}\label{sec0}
 
 Impedance provides a useful tool for solving dynamic problems in acoustics and elasticity.  A single scalar impedance is usually sufficient in 
acoustics, whereas a matrix of impedance elements is required to handle the vector nature of elastic wave motion, particularly in the presence of 
anisotropy.  The use of impedance matrices can offer new insight because their 
properties are intimately related to the  fundamental physics of the problem, as, for instance,   
the Hermitian property of the impedance matrix which is directly linked to  energy considerations. 
A classical example   is Lothe  \& Barnett's surface impedance matrix \cite{Lothe76,Barnett85} which
proved to be crucial for understanding surface waves in anisotropic 
homogeneous half-spaces, with the result that it provides perhaps the
simplest method for finding the Rayleigh wave speed. Biryukov  \cite{Biryukov85,Biryukov95}  has
developed a general impedance approach for surface waves in  inhomogeneous half-spaces  based on the differential Riccati
equation, see also \cite{Caviglia02}.  
Direct use of the impedance rather than the full displacement-traction wave field  provides  an efficient and stable procedure for computing high-frequency dispersion spectra.  Several numerical
schemes for guided waves and scattering in multilayered structures have been developed on this basis, e.g. \cite{Honein91,Wang02,Hosten03}. 
These involve the 3$\times $3 impedance matrix  often called the `surface impedance', although it actually differs from the 3$\times $3 surface
impedance of a half-space.  It is useful to  further  distinguish the familiar 3$\times $3 (`conditional') impedance from a   6$\times $6 (`two-point')  matrix  more closely related to the matricant of the system equations.   The nature  of these impedances have been analyzed using the Stroh  framework for homogeneous and functionally graded plates  \cite{Shuvalov00a,Shuvalov04a}.  
It is noteworthy that  both impedances are Hermitian
under appropriate physical assumptions; however, their Hermiticity implies a
somewhat different energy-flux property than the Hermiticity of the
half-space impedance. A bibliography on the impedance matrices for
piezoelectric media may be found in \cite{Biryukov95,Shuvalov08}.
 
The  above review concerns rectangularly anisotropic materials and planar
structures.  The objective of this paper is to provide an equally
comprehensive impedance formalism for time-harmonic modes of $n$th azimuthal
order in radially inhomogeneous cylindrically anisotropic materials of
infinite axial extent and various circular configurations. An important 
element in this task  is the Stroh-like state-vector formalism developed
for such materials in \cite{Shuvalov03}.  The results of \cite{Shuvalov03}, which
are based on the matricant in a Peano-series form (particularly the
definition of the `two-point' impedances similar to the case of planar
structures) are however only relevant  to a cylindrical annulus with no material
around the central point $r=0.$  The intrinsic singularity of elastodynamic
solutions at the origin  of the cylindrical coordinate system, which
rules out the Peano series, is an essential distinguishing feature as compared to the   
Cartesian setup.  The problem can be readily handled in (transversely)
isotropic homogeneous media with  explicit Bessel solutions;
however, it becomes considerably more intricate for cylindrically
anisotropic and for radially inhomogeneous solid cylinders. The main analytical
tool in  this case is the Frobenius series solution. The milestone results on
its application to homogeneous and layered cylinders of various classes of
cylindrical anisotropy include 
\cite{Mirsky66,Ohnabe70,Chou72,Srinivas74,Markus95,Martin01} 
(see also the review \cite{Soldatos94}). 
The state-vector formalism based on the Frobenius solution for
the general case of unrestricted cylindrical anisotropy and arbitrary radial
variation of material properties \cite{Shuvalov03a} is  of crucial importance to the 
present study. Another vital ingredient is the differential matrix
Riccati equation for an impedance \cite{Biryukov95}. To the best of the authors'
knowledge, this equation has only recently been used for the first time in
elasticity of cylinders by  \citet{Destrade09} who numerically solved it for
an elastostatic problem in tubes.
 
The presence of the special point $r=0$ distinguishes the solid-cylinder case   
from its Cartesian counterpart in many ways. Apart from the usual radiation 
condition at infinity, a similar kind of condition has to be applied at $r=0$.  
The   Riccati equation simultaneously   determines the central-impedance  at $r=0$ in a consistent manner  while  requiring it as the initial value for obtaining the
solid-cylinder impedance.  No other auxiliary 
(boundary) condition applies at $r=0,$ which is (again) unlike the
surface or conditional impedance for, say, a traction-free plane $y=0$.  These observations point to the fundamental role of the impedance formalism in cylindrically anisotropic
elastodynamics and actually call for a new type of the impedance matrix
appropriate for solid cylinders. The concept, properties and calculation of
the solid-cylinder impedance are among the main results of this paper.   


The outline  is as follows.   
Background material on the matricant, impedance matrices  and Riccati equations is  presented in \S\ref{sec1}  in a general context  not specific to cylindrical configurations.   In \S\ref{sec2} the governing equations 
for cylindrically anisotropic elastic solids are reviewed and  the first order differential system for the displacement-traction vector is  described.  Some examples of the use of impedance matrices are discussed in \S\ref{sec3}, and in the process the solid cylinder and the radiation impedance matrices are introduced.  Methods for determining the solid cylinder impedance are developed in \S\ref{sec4}.  This section  provides a detailed description of the Frobenius solution and its properties, and also discusses the Riccati solution.  Both methods involve the crucial central impedance matrix, to which  \S\ref{sec5} is devoted, where explicit solutions are presented and general attributes delineated, including the important Hermitian property.  The radiation impedance matrix is analyzed in \S\ref{sec6}.  Explicit examples are presented in \S\ref{sec7}  for the central impedance matrix in different types of anisotropy, and the solid cylinder  impedance is  explicitly presented  for transverse isotropy.  Numerical results illustrating the Riccati equation solution method are also discussed. Concluding remarks can be found in \S\ref{sec8}.

 \section{The matricant, impedance matrices and Riccati equations}\label{sec1}

 For the moment the development is  independent of the physical  dimension and the underlying coordinates.  
 Consider a  system of $2\nd$ linear ordinary differential equations 
\begin{equation}
 \frac{\dd \pmb{\eta} }{\dd y} =
\mathbf{Q}\pmb{\eta} \ \quad \mathrm{with}\ \mathbf{Q=}\left( 
\begin{array}{cc}
\mathbf{Q}_{1} & \mathbf{Q}_{2} \\ 
\mathbf{Q}_{3} & \mathbf{Q}_{4}%
\end{array}%
\right) ,\ \pmb{\eta =}\left( 
\begin{array}{c}
\mathbf{U} \\ 
\mathbf{V}%
\end{array}%
\right) .  \label{01}
\end{equation}%
The $\nd-$dimensional vectors $\mathbf{U}$, $\mathbf{V}$   and the $\nd\times \nd$ submatrices $\mathbf{Q}_j$, $j=1,2,3,4$ all  possess  uni-dimensional spatial dependence on $y$, which may be a Cartesian or radial coordinate.      The system matrix $\mathbf{Q}$ displays  an important algebraic symmetry which is a consequence of a  general flux continuity condition. The  derivative of the scalar quantity 
$ \pmb{\eta }^+  \mathbf{T}\pmb{\eta }$, where superscript `+' means the adjoint (complex conjugate transpose) and $\mathbf T$ has  block structure  with zero submatrices on the diagonal and  off-diagonal $\nd\times \nd$ identity matrices,  can be identified with the divergence of the flux vector $\mathbf P$ (to be defined more specifically later).  Thus, 
$ \frac{\dd }{\dd y}\big( \pmb{\eta }^+ \mathbf{T}\pmb{\eta }\big) \sim  \div {\mathbf P}$, and hence, \eqref{01} implies the 
connection between flux continuity and symmetry of the system matrix \cite{Shuvalov04a}
\begin{equation}
\mathbf{Q}=-\mathbf{TQ}^+\mathbf{T  }\quad \Leftrightarrow \quad
\div {\mathbf P} = 0 .  \label{91}
\end{equation}
The vanishing of $\div {\mathbf P} = 0 $ assumes certain physical restrictions
which will be described   when the elasticity problem is considered in Section \ref{sec2}.   

The $2\nd \times 2\nd$  matricant $\mathbf{M}  ( y,y_0 )$ is a function of two coordinates   defined as 
the solution of the initial value problem 
\beq{32}
  \frac{\dd \mathbf{M} }{\dd y}  
  ( y,y_0 )=\mathbf{Q}(y) \mathbf{M} ( y,y_0 ),
 \qquad  \mathbf{M} ( y_0,y_0) = \mathbf{I}_{(2\nd )} .
 \eeq
 The matricant  may be represented formally as a   Volterra or multiplicative integral evaluated by means of a Peano series \cite{Pease}, alternatively  it may be expanded in a Frobenius series
 \cite{Coddington}. 
 Let $\pmb{\eta}_\alpha(y)$  $(\alpha = 1,2,\ldots , 2\nd)$ be a set of partial solutions, that is, a complete set of independent solutions of the homogeneous system \eqref{01}, then 
$
 \mathbf{M} ( y,y_0 ) = \pmb{\mathcal{N}}(y) \pmb{\mathcal{N}}^{-1}(y_0)$,
 where $\pmb{\mathcal{N}}$ is the integral matrix (a first-rank tensor) $\pmb{\mathcal{N}}%
\left( y\right) =\left\Vert \pmb{\eta }_{1},...,\pmb{\eta }_{2\nd}\right\Vert $.   
 The propagator nature of the matricant is apparent from the property $
 \mathbf{M}  ( y,y_1 )\mathbf{M}  ( y_1,y_0 )= \mathbf{M}  ( y,y_0 )$, and in particular $\mathbf{M}  ( y,y_0 )
 = \mathbf{M}  ( y_0,y )^{-1}$, while the  symmetry \eqref{91}$_{1}$ implies 
 \beq{45}
      \mathbf{M} (y,y_0) = \mathbf{T}\mathbf{M}^+( y_0,y) \mathbf{T} . 
 \eeq
Hence,  
 \beq{45-}
      \mathbf{M}^{-1} (y,y_0) = \mathbf{T}\mathbf{M}^+( y,y_0) \mathbf{T} , 
 \eeq
 that is, $\mathbf{M}$    is  $\mathbf{T}$-unitary \cite{Pease}. 
    

In solving  problems one is often not  interested in the individual fields ${\mathbf U}(y)$ and ${\mathbf V}(y)$, but rather in  their relationship to one another, and perhaps only at one or two positions such as boundary values of $y$.  Accordingly we  introduce the 
 $\nd\times \nd$ conditional  impedance matrix $\z$ defined 
  such that  
\beq{00}
{\mathbf V} (y) =- i \z (y) {\mathbf U}(y) .
\eeq
 The  conditional nature of this impedance arises from an auxiliary condition at another coordinate $y_0$ \cite{Shuvalov00a,Shuvalov04a}, and may be understood  from an equivalent definition of the matricant  
\beq{023}
\begin{pmatrix}
\mathbf{U}\left( y\right) \\ 
\mathbf{V}\left( y\right)%
\end{pmatrix}%
  =
  \begin{pmatrix}
\mathbf{M}_{1} & \mathbf{M}_{2} \\ 
\mathbf{M}_{3} & \mathbf{M}_{4}%
\end{pmatrix}
\begin{pmatrix}
\mathbf{U}\left( y_0\right) \\ 
 \mathbf{V}\left( y_0\right) 
\end{pmatrix}, 
\quad
\text{where  }
\mathbf{M}( y,y_0) =  \begin{pmatrix}
\mathbf{M}_{1} & \mathbf{M}_{2} \\ 
\mathbf{M}_{3} & \mathbf{M}_{4}%
\end{pmatrix}
.
\eeq
Now suppose  $\z (y_0)$ is the conditional impedance at $y_0$, 
then 
\beq{045}
\begin{split}
&\mathbf{U}\left( y \right) =\left( \mathbf{M}_{1}-i\mathbf{M}_{2}\z (y_0) \right) \mathbf{U}\left( y_0\right) ,\\  
&\mathbf{V}\left( y\right) =\left( \mathbf{M}_{3}-i\mathbf{M}_{4}\z (y_0) \right) \mathbf{U}\left( y_0\right) ,%
\end{split}%
\nonumber
\eeq
and the conditional impedance at $y$ is therefore
\beq{46}
\z (y) = i\big( \mathbf{M}_{3}-i\mathbf{M}_{4}\z (y_0)  \big) \big( \mathbf{M}_{1}-i\mathbf{M}_{2}\z (y_0) \big)^{-1} .
\eeq
In practice, $\z (y_0)$ is often associated with   boundary conditions on the level surface $y=y_0$.  For instance, `zero traction'  and `rigid boundary' conditions are specified by vanishing $\mathbf V$ and $\mathbf U$, respectively, with  conditional impedances 
 \beq{047}
\z (y) = \begin{cases}
i  \mathbf{M}_{3}  \mathbf{M}_{1}^{-1}  & \text{zero traction }(\mathbf{V}\left( y_0 \right) ={\mathbf{0}} ),
\\
i  \mathbf{M}_{4}  \mathbf{M}_{2}^{-1}  & \text{rigid boundary }(\mathbf{U}\left( y_0 \right) ={\mathbf{0}} ) , 
\end{cases}
\eeq
 where $\mathbf{M}_{j}=\mathbf{M}_{j}\left( y,y_0\right) $ in eqs.\ \eqref{023}-\eqref{047}.

While it is possible to define  the conditional impedance  in terms of  solutions of the $2\nd \times 2\nd$ linear  system \eqref{01},  the same system leads through  a process of elimination  to a    
 quadratically nonlinear  equation for the $\nd \times \nd$ matrix $\z$: the differential Riccati equation \cite{Biryukov95}
\begin{equation}
\frac{\dd \z}{\dd y} +\z {\mathbf Q}_{1}-\mathbf{Q}_{4}\z -i\z \mathbf{Q}%
_{2}\z-i\mathbf{Q}_{3}=\mathbf{0}.  \label{2}
\end{equation}%
In this context the auxiliary impedance $\z (y_0)$ serves as an initial condition at $y=y_0$ which once specified uniquely determines $\z (y)$ at other positions.  
The symmetry \eqref{91}$_{1}$ renders eq. \eqref{2}
self-adjoint in the sense that if $\z$ is a solution then so is $\z^{+} $, which does not   imply  their equality.  It does however imply that the 
  differential Riccati equation \eqref{2}   produces an Hermitian impedance, 
   $\z (r)= \z^{+}(r)$, 
 as long as the initial condition is   Hermitian, $\mathbf{z}(y_0) =\mathbf{z}^+(y_0)$. We will also find useful the algebraic Riccati equation associated with eq.\ \eqref{2}, 
\begin{equation}
\z {\mathbf Q}_{1}-\mathbf{Q}_{4}\z -i\z \mathbf{Q}%
_{2}\z-i\mathbf{Q}_{3}=\mathbf{0},  \label{--2}
\end{equation}
the solution of which determines limiting values of the impedance, e.g. as $|y|\rightarrow \infty$, and can serve as the initial value for the differential equation \eqref{2}. 

We also introduce a `two-point' impedance $\mathbf{Z}\left( y,y_{0}\right)$ distinguished from the conditional impedance 
by its explicit 
dependence   upon   two arguments, and defined such that it relates 
the constituent parts of the $2\nd$-vector at $y$ and $y_0$ according to \cite{Shuvalov00a,Shuvalov04a}
\begin{equation}
\left( 
\begin{array}{c}
\mathbf{V}\left( y_{0}\right) \\ 
-\mathbf{V}\left( y\right)%
\end{array}%
\right) =-i\mathbf{Z}\left( y,y_{0}\right) \left( 
\begin{array}{c}
\mathbf{U}\left( y_{0}\right) \\ 
\mathbf{U}\left( y\right)%
\end{array}%
\right) .  \label{3}
\end{equation}%

Comparing \eqref{00} and \eqref{3} one might be tempted to surmise that 
the two-point impedance 
is  composed simply of block diagonal elements   $\mathbf{Z}_1 (y,y_0)$ and $\mathbf{Z}_4(y,y_0)$ identified as $ \z ( y_0)$  and $ - \z ( y ) $, respectively,  where $\z $ is the conditional impedance,  and with zero off-diagonal blocks ($\mathbf{Z}_3 $ and $\mathbf{Z}_2$).  But the two-point impedance is more fundamental  and thereby richer, as one can see  by comparing  \eqref{023} and \eqref{3},  implying 
\bal{96}
\begin{pmatrix}
\mathbf{Z}_{1} & \mathbf{Z}_{2} \\ & \\
\mathbf{Z}_{3} & \mathbf{Z}_{4}%
\end{pmatrix} 
&=i  
\begin{pmatrix}
- \mathbf{M}_{2}^{-1}\mathbf{M}_{1} & \mathbf{M}_{2}^{-1}  
\\ & \\
\mathbf{M}_{4}\mathbf{M}_{2}^{-1}\mathbf{M}_{1} -\mathbf{M}_{3}   &  \quad -\mathbf{M}_{4} \mathbf{M}_{2}^{-1}
\end{pmatrix} , &
  &\det \mathbf{Z} = \frac{\det (-\mathbf{M}_3) }{\det \mathbf{M}_2 }, 
\nonumber \\  & \\ 
\begin{pmatrix}
\mathbf{M}_{1} & \mathbf{M}_{2} \\ & \\
\mathbf{M}_{3} & \mathbf{M}_{4}%
\end{pmatrix} 
&=  
\begin{pmatrix}
- \mathbf{Z}_{2}^{-1}\mathbf{Z}_{1} & i\mathbf{Z}_{2}^{-1}  
\\ & \\
i \mathbf{Z}_{3} -i\mathbf{Z}_{4}\mathbf{Z}_{2}^{-1}\mathbf{Z}_{1}    &  \quad -\mathbf{Z}_{4} \mathbf{Z}_{2}^{-1}
\end{pmatrix} ,  
&
 &\det \mathbf{M} = \frac{\det \mathbf{Z}_3 }{\det \mathbf{Z}_2 }, 
\nonumber  
\end{align}
where $\mathbf{M}_{j}=\mathbf{M}_{j}\left( y,y_0\right) $, $\mathbf{Z}_{j}=\mathbf{Z}_{j}\left( y,y_0\right) $.
The identity \eqref{45}  then implies  the important properties that the   two-point impedance is Hermitian, and that the matricant determinant is of unit magnitude, i.e., 
\beq{356}
\text{Eq. }\eqref{91}_1 \quad \Rightarrow \quad
\mathbf{Z} =\mathbf{Z}^+, 
\qquad 
\det {\mathbf M} = e^{i \phi } \quad\text{where } \phi = \arg \det ({\mathbf M}_1{\mathbf M}_4) .   
\eeq
It   follows directly from $\eqref{32}_1$ and Jacobi's formula that the phase satisfies the differential equation
$\frac{\dd \phi}{\dd y} = -  i \tr \mathbf{Q} $ with initial condition $\phi (y_0)=0$. The matricant is therefore unimodular 
$(\det {\mathbf M}=1)$ if  $\tr \mathbf{Q} $ vanishes. 
   Further properties of the impedance may be deduced by swapping the `running' and
'reference' points $y$ and $y_{0}$ in \eqref{3} (i.e. in both $\mathbf{U}$, $%
\mathbf{V}$ and $\mathbf{Z}$), implying the reciprocal form
\begin{equation}
\left( 
\begin{array}{c}
\mathbf{V}\left( y\right) \\ 
-\mathbf{V}\left( y_{0}\right)%
\end{array}%
\right) =-i\mathbf{Z}\left( y_{0},y\right) \left( 
\begin{array}{c}
\mathbf{U}\left( y\right) \\ 
\mathbf{U}\left( y_{0}\right)%
\end{array}%
\right) ,  \label{4} \nonumber
\end{equation}%
whence follows an obvious relation 
\begin{equation}
\mathbf{Z}\left( y_{0},y\right) =-\mathbf{TZ}\left( y,y_{0}\right) \mathbf{T}.   \label{-5}
\end{equation}
The two-point impedance therefore has the structure
\beq{012}
{\mathbf Z}(y,y_0) = 
\begin{pmatrix}
{\mathbf Z}_1   & {\mathbf Z}_2 
\\  & \\ 
  {\mathbf Z}_3 &  {\mathbf Z}_4 
\end{pmatrix} , 
\quad \text{with} \quad
\begin{matrix}
{\mathbf Z}_1(y,y_0) = -  {\mathbf Z}_4(y_0,y), 
\\
{\mathbf Z}_2(y,y_0) = -  {\mathbf Z}_3(y_0,y) . 
\end{matrix}
\eeq
As an alternative to eq.\ \eqref{46} 
the  conditional impedance at $y$  may be  expressed in terms of the impedance at $y_0$ by using  the two-point impedance,
 \beq{146}
 \z (y ) = -\mathbf{Z}_{4}-\mathbf{Z}_{3}\big( \z (y_0)-\mathbf{Z}_{1}\big)^{-1}\mathbf{Z}_{2},
 \eeq
where  $\mathbf{Z}_{j}=\mathbf{Z}_{j}\left( y,y_0\right) $.  Note that $ \z (y )$ is Hermitian if $ \z (y_0 )$ is. 

 In the same way that the matricant $\mathbf{M}(y,y_0) $ satisfies an ordinary differential equation in $y$, viz. $\eqref{32}$, it is possible to express the dependence of  $\mathbf{Z} (y,y_0) $ on $y$ in differential form.  
Differentiating \eqref{3} with respect to $y$ and using \eqref{01} to eliminate the traction vectors yields an  equation for the two-point impedance, 
\begin{equation}
\frac{\dd \mathbf{Z}}{\dd y}  +\mathbf{ZJ}_{1} - \mathbf{J}_{4}%
\mathbf{Z} +i\mathbf{ZJ}_{2}\mathbf{Z}+i\mathbf{J}_{3}=\mathbf{0,\ }\mathrm{%
where\ }\mathbf{J}_{j}=\left( 
\begin{array}{cc}
\mathbf{0} & \mathbf{0} \\ 
\mathbf{0} & \mathbf{Q}_{j}(y)%
\end{array}%
\right) .  \label{813}
\end{equation}%
The self-adjoint property of this  equation is obvious because the  two-point impedance is itself self-adjoint
(Hermitian).   Direct integration of the  differential system \eqref{813} subject to initial conditions at $y=y_0$ is problematic because of the fact that all   submatrices of $\mathbf{Z} (y,y_0) $ are of the form $\pm i (\int_{y_0}^y\dd y \, {\mathbf Q}_2)^{-1}$ as $|y-y_0|\rightarrow 0$, and hence undefined.   Differential equations with well defined (finite) initial value conditions  can be obtained  for the block matrices $\mathbf{Z}_j^{-1}$  $(j=1,2,3,4)$ by simple manipulation of eq.\ \eqref{813}, but we do not discuss this further here. It is interesting to note, however, that  
inspection of the block structure of \eqref{813} shows that the equation for $\mathbf{Z}_4 $ decouples from the other submatrices and it is the same as  the differential Riccati equation \eqref{2} for the conditional impedance (under the interchange $
{\mathbf Z}_4 \leftrightarrow -\z$).   Furthermore, since $\mathbf{Z}_4$  becomes unbounded as $y\rightarrow y_0$, eq.\ \eqref{813} implies that the submatrix  $-\mathbf{Z}_4$ is the conditional impedance with the auxiliary condition of  rigid (infinite) impedance at $y=y_0$, an observation that is verified by   eqs.\ \eqref{047}$_2$ and   \eqref{96}$_1$. 

 \section{Cylindrically anisotropic elastic solids}\label{sec2}

 \subsection{Equations in cylindrical coordinates}
 
 The   dynamic equilibrium equations  for a linearly elastic material when expressed in cylindrical coordinates are  \cite{Ting96a} 
\beq{5}
r^{-1}(r\mathbf{t}_r)_{,\,r}+r^{-1} (\mathbf{t}_{\theta,\,
\theta}+\mathbf{K}\,\mathbf{t}_{\theta})+\mathbf{t}_{z,\,z}=   \rho
\ddot{\mathbf u} 
\quad\text{with  }
 \mathbf K\,=
\begin{pmatrix}
{0} & -1 &0
\\
1 & 0 & 0
\\
0 & 0 &0
\end{pmatrix}.
 \eeq
 Here $\rho=\rho ({\mathbf x})$ is the mass density, ${\mathbf u} ={\mathbf u} ({\mathbf x} ,t)$ the displacement, and the traction vectors 
 $\mathbf{t}_i=\mathbf{t}_i ({\mathbf x}, t)$, $i=r,\theta,z$,  are defined by the orthonormal basis vectors 
 $\{ \mathbf{e}_r, \mathbf{e}_\theta, \mathbf{e}_z\}$ 
 of the cylindrical coordinates 
 $\{r, \theta ,z\}$ according to 
 $\mathbf{t}_i   = \mathbf{e}_i \pmb{\sigma}$ $(i=r,\theta,z)$,  
 where $  \pmb{\sigma} ({\mathbf x} ,t)$ is the stress, 
and a comma denotes partial differentiation. 
 With  the same basis vectors,  and assuming the summation convention on repeated indices,  the 
elements of stress are $\sigma_{ij} = c_{ijkl} \varepsilon_{kl}$ where 
$\pmb{\varepsilon} = \frac12 (\nabla {\mathbf u} + \nabla {\mathbf u}^T)$ is the strain, 
$ c_{ijkl} = c_{ijkl} ({\mathbf x})$ are elements of the fourth order  (anisotropic) elastic stiffness tensor,  
and $T$ denotes transpose.    The 
traction vectors are    
\cite{Shuvalov03}
 \beq{-1}
\begin{pmatrix}
 \mathbf{t}_r 
 \\   \\
  \mathbf{t}_{\theta}
  \\  \\
  \mathbf{t}_z  
 \end{pmatrix}
 =
 \begin{pmatrix}
  \mathbf{ \widehat{ Q}} & \mathbf{R} & \mathbf{P}
  \\ & & \\
   \mathbf{R}^T &  \mathbf{ \widehat{ T}} & \mathbf{S}
 \\ & & \\
  \mathbf{P}^T &  \mathbf{S }^T &\mathbf{ \widehat{ M}}
 \end{pmatrix}
 \begin{pmatrix}
 \mathbf{u}_{,\,r}
 \\  \\
  \frac1r(\mathbf{u},_{\,\theta}+\mathbf{K}\,\mathbf{u})
  \\  \\
  \mathbf{u}_{,\,z}
 \end{pmatrix},
 \quad
\begin{pmatrix}
  \mathbf{ \widehat{ Q}} & \mathbf{R} & \mathbf{P}
  \\ & & \\
   \mathbf{R}^T &  \mathbf{ \widehat{ T}} & \mathbf{S}
 \\ & & \\
  \mathbf{P}^T &  \mathbf{S }^T &\mathbf{ \widehat{ M}}
 \end{pmatrix}
=
 \begin{pmatrix}
   (e_r e_r) &  (e_r e_\theta ) &  (e_r e_z )
  \\ & & \\
   &   (e_\theta e_\theta ) &  (e_\theta e_z )
 \\ & & \\
   &    &  (e_z e_z )
 \end{pmatrix}, \nonumber
 \eeq
where, in   the notation of \cite{Lothe76},    the matrix $\left( ab\right) $ has 
components $\left( ab\right) _{jk}=a_{i}c_{ijkl}b_{l}$  for  arbitrary vectors $\mathbf{a}$ and
$\mathbf{b}$.
The explicit form  of the various   matrices is apparent with  the use of  
Voigt's notation $c_{ijkl}\rightarrow c_{\alpha\beta}$ $(\alpha, \beta \in \{ 1,2,\ldots , 6\})$ 
\bal{783}
\mathbf{\widehat{Q}}=
\begin{pmatrix}
c_{11}&c_{16}&c_{15}
\\
c_{16}&c_{66}&c_{56}
\\
c_{15}&c_{56}&c_{55}
\end{pmatrix},
\quad 
&
\mathbf{\widehat{T}}=
\begin{pmatrix}
c_{66}&c_{26}&c_{46}
\\
c_{26}&c_{22}&c_{24}
\\
c_{46}&c_{24}&c_{44}
\end{pmatrix},
\quad
\mathbf{\widehat{M}}= 
\begin{pmatrix}
c_{55}&c_{45}&c_{35}
\\
c_{45}&c_{44}&c_{34}
\\
c_{35}&c_{34}&c_{33}
\end{pmatrix},
\nonumber \\
& \\ \nonumber
\mathbf{S}=
\begin{pmatrix}
c_{56}&c_{46}&c_{36}
\\
c_{25}&c_{24}&c_{23}
\\
c_{45}&c_{44}&c_{34}
\end{pmatrix},
\quad 
&
\mathbf{P}=
\begin{pmatrix}
c_{15}&c_{14}&c_{13}
\\
c_{56}&c_{46}&c_{36}
\\
c_{55}&c_{45}&c_{35}
\end{pmatrix},
\quad 
 \mathbf{R}=
\begin{pmatrix}
c_{16}&c_{12}&c_{14}
\\
c_{66}&c_{26}&c_{46}
\\
c_{56}&c_{25}&c_{45}
\end{pmatrix} .
\nonumber
\end{align}

 \subsection{Cylindrically anisotropic materials}

The concept of cylindrical anisotropy, which apparently originated with Jean Claude Saint-Venant,    and has been   elaborated by   \citet{Lekhnitskii},   demands  the angular independence of material constants in the cylindrical coordinates, but admits    their
dependence on $r$ and as well on $z$.   We consider  materials with no axial dependence whose 
 density and the elasticity tensor  may depend upon    $r$, i.e.
$\rho=\rho(r)$ and $c_{ijkl}=c_{ijkl}(r)$   $\forall\,
i,j,k,l\in {r,\theta, z}$.
We   seek  solutions  in the form of time-harmonic cylindrical
waves as 
 \beq{10}
 \mathbf u=\mathbf{U}^{(n)}(r)   \ee^{i (n\theta  +k_z z - \omega t) }, \quad
 \mathbf{t}_r=\mathbf{\Upsilon}^{(n)}(r) \ee^{i (n\theta  +k_z z - \omega t) },
 \eeq
 where  $n=0,1,2,\ldots $ is the circumferential number. 
 
 The dependence of the displacement and traction on the single spatial coordinate $r$ allows the elastodynamic equations to be reduced to the canonical form of eq.\ \eqref{01} \cite{Shuvalov03}: 
\beq{130}
\frac{\dd}{\dd r}\pmb{\eta}^{(n)}(r)=    \frac{i}{r}\,\mathbf{G}(r)\pmb{\eta}^{(n)}(r) ,
 \eeq
where $\pmb{\eta}^{(n)}$ is a $6\times 1$ vector 
\beq{14}
\pmb{\eta}^{(n)}(r)= 
\begin{pmatrix}
{\mathbf{U}^{(n)}(r)}
\\
{\mathbf{V}^{(n)}(r)}
\end{pmatrix},\quad \text{with  } {\mathbf{V}^{(n)}(r)} = {ir\mathbf{\Upsilon}^{(n)}(r)},
\eeq
 and the $6\times 6$ system matrix $\mathbf{G}$ is defined by 
 \beq{15}
i\mathbf{G}(r)=\mathbf g_0(r)+r\mathbf g_1(r)+r^2\mathbf g_2(r)
= \begin{pmatrix} {\mathbf g^{\{1\}}}(r) &{i\mathbf g^{\{2\}}}(r)
\\
{i\mathbf g^{\{3\}}}(r) &{-\mathbf g^{\{1\}+}}(r)
\end{pmatrix}. \nonumber
 \eeq
 The individual $6\times 6$   matrices are  
\beq{160}
 \mathbf g_0=\begin{pmatrix} {\mathbf g_0^{\{1\}}}&{i\mathbf g_0^{\{2\}}}
\\
{i\mathbf g_0^{\{3\}}}&{-\mathbf g_0^{\{1\}+}}
\end{pmatrix},\quad
 \mathbf g_1=ik_z\begin{pmatrix} {\mathbf g_1^{\{1\}}}&{\mathbf{0}}
\\
{i\mathbf g_1^{\{3\}}}&{\mathbf g_1^{\{1\}T}}
\end{pmatrix},\quad
 \mathbf g_2=\begin{pmatrix} {\mathbf{0}}&{\mathbf{0}}
\\
{i\mathbf g_2^{\{3\}}}&{\mathbf{0}}
\end{pmatrix},\nonumber
\eeq
with the   $3\times 3$ matrices
\beq{163}
 \mathbf g^{\{1\}}=\mathbf g_0^{\{1\}} + ik_z r \mathbf g_1^{\{1\}},
 \quad
 \mathbf g^{\{2\}}=\mathbf g_0^{\{2\}},
 \quad
 \mathbf g^{\{3\}}=\mathbf g_0^{\{3\}} + ik_z r \mathbf g_1^{\{3\}} +r^2 \mathbf g_2^{\{3\}} .
\nonumber 
 \eeq
The constituent $3\times 3$ matrices are 
\bal{16}
 \mathbf g_0^{\{1\}}& =- \mathbf{\widehat{Q}}^{-1}\widetilde{\mathbf R} ,&  
 \mathbf g_1^{\{1\}}&=-\mathbf{\widehat{Q}}^{-1}\mathbf{P},
 \nonumber 
 \\
 \mathbf g_0^{\{2\}}&=-\mathbf{\widehat{Q}}^{-1}=\mathbf g_0^{\{2\}T},& 
 \mathbf g_1^{\{3\}}&=    \mathbf{P}^T \mathbf{\widehat{Q}}^{-1} \widetilde{\mathbf R} -\widetilde{\mathbf S}
-\big( \mathbf{P}^T \mathbf{\widehat{Q}}^{-1} \widetilde{\mathbf R} -\widetilde{\mathbf S} \big)^+
 =-\mathbf
 g_1^{\{3\}+}, \quad
 \nonumber  \\
 \mathbf g_0^{\{3\}}&= \widetilde{\mathbf T} - \widetilde{\mathbf R}^+ \mathbf{\widehat{Q}}^{-1}\widetilde{\mathbf R} =\mathbf
 g_0^{\{3\}+},&
\mathbf g_2^{\{3\}}&=k_z^2(\widehat{\mathbf
M}-\mathbf{P}^T\mathbf{\widehat{Q}}^{-1}\mathbf{P})-\rho\omega^2 \mathbf I=\mathbf{g}_2^{\{3\}T}, 
\nonumber 
\end{align}
 where 
 \beq{1122}
\widetilde{\mathbf R} = {\mathbf R}\pmb{\kappa},
\qquad
\widetilde{\mathbf S} = \pmb{\kappa}{\mathbf S},
\qquad
 \widetilde{\mathbf T} = \pmb{\kappa}^+\widehat{\mathbf T}\pmb{\kappa} = \widetilde{\mathbf T}^+ ,
 \qquad
 \pmb{\kappa}=\mathbf{K}+ in\mathbf{I} =-\pmb{\kappa}^+ .
 \nonumber
\eeq
 The matrices $\mathbf{g}_0^{ \{2\} }$ 
 and $\mathbf{g}_0^{ \{3\}}$ are negative definite and positive semi-definite, respectively,  for real-valued and positive definite elastic moduli. 
Note that the $n$th order modal solution  $\pmb{\eta}^{(n)}(r) $ is a function of the radial coordinate, but it is also an implicit function of the frequency $\omega$ and the axial wavenumber  $k_{z}$, which  dependence is 
here kept tacit.  In the same manner, the dependence of $\mathbf{G}(r) $  upon $n$, $\omega$ and $k_z$ is understood.  

The superscript $^{(n)}$ is  omitted henceforth, with the exception of the specific cases $n=0$, $1$, as required. 


\subsection{Cylindrical elasticity in the general context}

The results of the previous subsection, particularly eqs.\ \eqref{130} and \eqref{14}, show that the cylindrically anisotropic system of azimuthal order $n$ is a special case of the  formulation of Section \ref{sec1} generally with $\nd = 3$, and $\{ y,  \mathbf{U}, \mathbf{V}, \mathbf{Q} \} \rightarrow  \{ r,\, \mathbf{U}\o\, , ir\mathbf{\Upsilon}\o,\,  ir^{-1}\mathbf{G} \}$.  The physical restrictions required for the Hermiticity condition \eqref{91} are real-valued $\omega$, $k_z$  and material constants (more precisely, Hermitian elastic moduli $c_{\alpha \beta} = c_{ \beta \alpha}^*$  suffice \cite{Shuvalov08}).  Under these conditions the 
  $6\times 6$ matrix $\mathbf{G}(r) $ displays  the symmetry  
 \beq{056}
\mathbf{G} =  \mathbf{T}\mathbf{G}^+\mathbf{T}.   
\eeq
The 6$\times $6 matricant $\mathbf{M}\o ( r,r_0 )$ is     
the solution of the initial value problem
\beq{132}
 \bigg( \frac{i}{r}\,\mathbf{G}(r)-\frac{\dd}{\dd r} \bigg) 
 \mathbf{M}\o( r,r_0 )=\mathbf{0} ,
 \qquad  \mathbf{M}\o( r_0,r_0) = \mathbf{I}_{(6)},\quad r ,  \, r_0 \ne 0.
 \eeq
 The condition that $r$ and $r_0$ are strictly positive is important since the case of zero radial coordinate  needs to be handled separately, which is discussed at length below.  Note that we do not specify whether $r$ or $r_0$ is the greater or lesser of the two radii. 
The matricant allows us to express the 
  state vector $\pmb{\eta }\o\left( r\right) $ of partial modes in a cylinder  as 
\beq{807}
\pmb{\eta }\o(r) =\mathbf{M}\o\left( r,r_0\right) \pmb{\eta }\o(r_0)
,\quad r ,  \, r_0 \ne 0.    
\eeq 

The pointwise elastodynamic energy balance   is ${\dd \cal E}/{\dd t} + \div {\cal P} = 0$ where 
${\cal  E} $  is the energy density per unit volume and  ${\cal P} $  the energy flux vector.  
  The pertinent form of $\eqref{91}_2$ for  cylindrical elasticity  is 
$\div {\mathbf P} =r^{-1}\dd (rP_r\o)/\dd r = 0$ where $P_r\o = \langle {\cal P}\rangle_t \cdot {\mathbf e}_r$ is the time averaged radial component for azimuthal mode $n$,
\beq{010}
P_r\o (r) = -\frac{\omega }{4r} \, {\pmb{\eta}\o}^+(r) \mathbf{T} \pmb{\eta}\o(r),  
\eeq
 which together with the system equation \eqref{130} implies the symmetry \eqref{056} for $\mathbf G$ (see also \cite{Shuvalov03}).   

The conditional impedance matrix $\mathbf{z}\o  $ relates   traction and displacement at a particular value of $r$, but specifically $r\ne 0$, according to eq.\ \eqref{00}. 
The point  $r=0$ requires a separate discussion, and indeed a newly defined impedance, introduced in the next section.  For the moment we note that $ \mathbf{z}\o (r)$ is contingent upon the definition of the (one-point) impedance at some  radial coordinate, say $ \mathbf{z}\o (r_0)
=  \mathbf{z}\o_0$.    The traction  at other values of $r$ is then unambiguously related to the local displacement  by either the matricant or the two-point impedance matrices, using equation  \eqref{46} or \eqref{146}.  By rewriting eq.\ \eqref{010} we see  
that the conditional impedance  determines  the pointwise flux,
\beq{-010}
P_r\o (r) = -\frac{\omega }{2r} \, \Im \, \big\{  \mathbf{U}\o^+ (r)  \mathbf{z}(r)
 \mathbf{U}\o (r) \big\},  
\eeq
which is zero for all $\mathbf{U}\o (r)$ only if $\mathbf{z}$ is Hermitian. 
This in turn is the case only if    $\mathbf{z}(r_0)=\mathbf{z}_0$ is Hermitian, i.e., if there is no flux across the surface $r=r_0$.  
On the other hand, the  6$\times $6 two-point impedance matrix  $\mathbf{Z}\o( r_2,r_1)$ of eq.\ \eqref{3} defines   the 
global energy flow into or out of  the finite  region between the two radial coordinates $r_1< r_2$. 
Let $E\o(t)$ be the total energy in the shell cross-section 
per unit length of the cylinder for azimuthal  mode $n$.  Its increment over one period of time harmonic motion is
\beq{2351}
\Delta E\o = - 2\pi \frac{2\pi}{\omega}  \left. \big( rP_r\big)\right|_{r_1}^{r_2} =-2
\pi^2  \Im \, \bigg\{
\begin{pmatrix}
\mathbf{U}\o(r_1) \\ 
\mathbf{U}\o(r_2)%
\end{pmatrix}^+  
 \mathbf{Z}\o( r_2,r_1) 
 \begin{pmatrix}
\mathbf{U}\o(r_1) \\ 
\mathbf{U}\o(r_2)%
\end{pmatrix} \bigg\},   
\eeq
which is identically zero for real  $\omega$, $k_{z}$ and Hermitian parameters, i.e., when $\mathbf{Z}\o$ is Hermitian.
  If the material in the slab is lossy  then 
$  \Im (\mathbf{Z}\o -\mathbf{Z}\o^+)  $ should be positive definite in order that $ E\o $ is not increasing with time. 


The differential Riccati equation satisfied by the one-point impedance matrix follows from   \eqref{2}
 as
 \beq{2221}
 r\frac{\dd {\mathbf z}}{\dd r}
 + {\mathbf z} {\mathbf g}^{\{1\}}
  + {\mathbf g}^{\{1\}+} {\mathbf z} 
 + {\mathbf z} {\mathbf g}^{\{2\}} {\mathbf z} 
  +{\mathbf g}^{\{3\}}
=  \mathbf{0}. 
\eeq
The initial value problem for $\mathbf{z}\o (r)$ is therefore
\beq{222}
 r\frac{\dd {\mathbf z}\o}{\dd r}
 -
 \big[  {\mathbf z}\o +\big(\widetilde{\mathbf R} + ik_z r {\mathbf P}\big)^+\big]
 {\mathbf{\widehat{Q}}}^{-1} \big[  {\mathbf z}\o + \widetilde{\mathbf R} + ik_z r {\mathbf P} \big]
 + {\mathbf B}(r)={\mathbf{0}}  , \quad r> 0, 
 \quad 
 \mathbf{z}\o (r_0)
=  \mathbf{z}\o_0,
\eeq
where
\beq{224}
 {\mathbf B}(r) =\widetilde{\mathbf T}
  + ik_z r ( \widetilde{\mathbf S}^+- \widetilde{\mathbf S}  ) + r^2( k_z^2 \widehat{\mathbf M} -
  \rho \omega^2  {\mathbf I}) =  {\mathbf B}^+ (r).
 \eeq
 Equation  \eqref{222}  shows the explicit dependence upon $\omega$, $k_z$ and the elastic moduli.   
The exclusion of the distinguished point $r=0$ at the cylinder centre  is addressed  next. 

\begin{figure}[htbp]
				\begin{center}	
				\includegraphics[width=3.6in , height=3in 					]{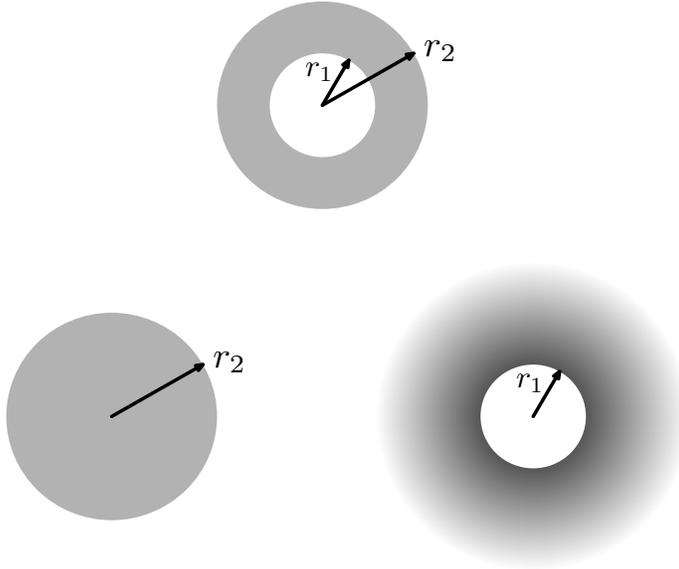} 
	\caption{Three types of cylindrical structures defined by $r_1 \le r\le r_2$: the annulus $(0<r_1 < r_2<\infty)$, the solid $(r_1=0)$, and the exterior region $(r_2=\infty)$. }
		\label{fig1} \end{center}  
	\end{figure}


\section{Wave impedance matrices for  cylinders}\label{sec3}

In this section we describe  typical uses of impedance matrices, and in  the process introduce the solid-cylinder impedance ${\mathbf Z}(r)$ and the   radiation impedance
${\mathbf Z}_{\mathrm {rad}}(r)$.  
We consider the three  distinct configurations depicted schematically in Figure  \ref{fig1}.

\subsection{Solid-cylinder impedance matrix ${\mathbf Z}\o(r)  $}

A solid cylinder, by definition, is one that  includes the axis $r=0$. 
A new impedance matrix 
 is introduced to handle this situation. 
The solid-cylinder impedance ${\mathbf Z}\o(r)  $  is defined in the usual manner by its property of relating  the traction and displacement $3-$vectors of \eqref{10}  according to
\beq{21}
 \mathbf{V}\o(r) = - i{\mathbf Z}\o(r) {\mathbf U}\o(r), \quad r \ge 0,
\eeq
although this  is   not  a conditional impedance matrix  because of the absence of an auxiliary  impedance condition at some other coordinate.  Instead, the solidity of the cylinder at $r=0$ dictates the 
character of  ${\mathbf Z}\o(r)$ (and one could argue that it is `conditional' in that sense). 
The limiting value of the solid impedance at $r=0$ plays a crucial role, and we accordingly define the central-impedance    matrix:
\beq{445}
\mathbf{Z}\o_0 \equiv \mathbf{Z}\o(0).
\eeq 
The  properties of the  central impedance  are discussed  in detail in \S\ref{sec5} after 
we develop  methods for finding the solid-cylinder impedance matrix in \S\ref{sec4}.

As an example application consider the task of finding the dispersion equation for guided waves of frequency $\omega$ and wavenumber $k_z$.  
We suppose, quite generally,    an interface condition  on
the level surface $r=r_2$ of the form   
\beq{3.89}
\mathbf{V }\o(r_2)  = -   i\mathbf{z}\o_2  
\mathbf{U}\o(r_2),    
\eeq
where $\mathbf{z}\o_2$ is considered as  given.  It could  be zero (traction free condition), infinite (rigid boundary), or it could be defined by some surrounding material, whether finite or infinite in extent.  For instance, if the solid cylinder is surrounded by a shell of cylindrically anisotropic material
 in lubricated contact at $r=r_2$ and free at $r=r_3>r_2$, then $\mathbf{z}\o_2 = z_{11}(r_2) \mathbf{e}_r\mathbf{e}_r^T$ where $\mathbf{z}(r_2)$ is the conditional impedance with the auxiliary condition $\mathbf{z}(r_3)
=\mathbf{ 0}$.   
Assuming \eqref{3.89} describes the   condition at the outer surface, the desired dispersion equation is  
\beq{-423} 
\det\big( {\mathbf Z}\o(r_2) -   \mathbf{z}\o_2 \big)=0. 
\eeq
It is instructive to compare \eqref{-423} with the dispersion equation for
a (possibly functionally graded) layer $y\in \left[ 0,y_{2}\right] $ with
the traction-free surface $y=0$ on a homogeneous substrate $y>y_{2}$ which
may be written in the form \cite{Shuvalov02}%
\begin{equation*}
\det \big( \mathbf{z}\left( y_{2}\right) -\mathbf{Z}_{2}\big) =0,
\nonumber 
\end{equation*}%
where $\mathbf{Z}_{2}$ is a (constant) impedance of the substrate and $%
\mathbf{z}\left( y_{2}\right) $ is the conditional impedance of the layer
satisfying the  reference  condition $\mathbf{z%
}\left( 0\right) =0.$  If the surrounding material
beyond a rigid (say) interface $r=r_{2}$ is infinite, then there are
Stoneley-like waves defined by the dispersion equation \eqref{3.89} with $\mathbf{z%
}_{2}=\mathbf{Z}_{\mathrm{rad}}\left( r_{2}\right) ,$ where $\mathbf{Z}_{%
\mathrm{rad}}\left( r\right) $ is the radiation impedance discussed in \S\ref{sec6}. 

The solid-cylinder impedance also provides a means to compute the modal displacement vector ${\mathbf U}(r)$ for all $0\le r \le r_2$ if the dispersion equation \eqref{-423} is satisfied.  By analogy with the case of an annulus \cite{Destrade09}, the unnormalized  displacement follows from the system equation \eqref{130} and the definition of ${\mathbf Z}\o(r)$ in \eqref{21} as the solution of the  initial value problem 
\beq{-72}
r\frac{\dd {\mathbf U}} {\dd r} +
\mathbf{\widehat{Q}}^{-1}  \big(\widetilde{\mathbf R}+ik_z {\mathbf P} +
 {\mathbf Z} \big){\mathbf U}(r)={\mathbf{0}}, \, \, \,
0\le r \le r_2; 
\quad {\mathbf U}(r_2) = {\mathbf U}_0, 
\eeq
where ${\mathbf U}_0$ is the null vector of the surface impedance condition, 
$\big( {\mathbf Z}\o(r_2) -   \mathbf{z}\o_2 \big){\mathbf U}_0 =0$. 
Note that the solution of \eqref{-72} remains well behaved  even as the matricant solution ${\mathbf U}(r)=\big( {\mathbf M}_1(r,r_2)-i{\mathbf M}_2(r,r_2){\mathbf Z}(r_2)\big) {\mathbf U}(r_2)$ is numerically unstable, see \S\ref{4.2.2}.

\subsection{Impedance matrices for cylinders of infinite radius}\label{sec3.2}

Consider  a cylinder extending to infinity in the radial direction, with inner surface at $r=r_1$, see Figure \ref{fig1}. 
A  wave incident from $r>r_1$  results in a total field that  can be expanded in terms of partial waves of the form \eqref{10}.  The amplitude of the $n^\text{th}$ azimuthal 
mode is 
\beq{-3}
{\mathbf U}\o(r) = {\mathbf U}\o_\mathrm{inc}(r)+{\mathbf U}\o_\mathrm{scat}(r), \quad r\ge r_1,
\eeq
where  
 the scattered amplitude ${\mathbf U}\o_\mathrm{scat}(r)$ satisfies a radiation condition  at $r \rightarrow \infty$.   This in turn requires that  the following condition prevails on the interface:
\beq{+6}
\mathbf{V }\o_\mathrm{scat}(r)  =    -i \mathbf{Z}\o_\mathrm{rad}(r)  
\mathbf{U}\o_\mathrm{scat}(r),  \quad r=r_1,
\eeq
where the radiation impedance matrix $\mathbf{Z}\o_\mathrm{rad}(r)$ is   defined by the radiation conditions, see \S\ref{sec6}.
The scattered field is then uniquely determined by the  condition at $r=r_1$, which we assume is of the generalized form \eqref{3.89} with prescribed interface impedance $\mathbf{z}\o_1$.  Then, 
\beq{-7}
- \mathbf{Z}\o_\mathrm{inc}(r_1)  {\mathbf U}\o_\mathrm{inc}(r_1) -\mathbf{Z}\o_\mathrm{rad}(r_1) {\mathbf U}\o_\mathrm{scat}(r_1)
= - \mathbf{z}\o_1 \big( {\mathbf U}\o_\mathrm{inc}(r_1)+{\mathbf U}\o_\mathrm{scat}(r_1)\big)
,   \nonumber
\eeq
where $\mathbf{Z}\o_\mathrm{inc}(r) $ is the impedance of the incident wave, which follows directly from the equations of motion.   The scattered amplitude on the interface is therefore
\beq{-8}
{\mathbf U}\o_\mathrm{scat}(r_1)
= \big( \mathbf{z}\o_1 - \mathbf{Z}\o_\mathrm{rad}(r_1)  \big)^{-1}
	\big(\mathbf{z}\o_1 - \mathbf{Z}\o_\mathrm{inc}(r_1)  \big) \, {\mathbf U}\o_\mathrm{inc}(r_1),
\eeq
  which  provides the initial condition to  determine  the entire scattered field in $r\ge r_1$.  Further details on the radiation impedance matrix are provided in \S\ref{sec6}, including its asymptotic properties for large $r$.

\subsection{An annulus of finite thickness}
The case of the annulus $0< r_1\le r\le r_2$ fits   readily into the general theory.  Again consider the task of finding the dispersion equation for guided waves, which may    be found by simultaneous satisfaction of the   conditions on the two radial surfaces.  Suppose the conditions are both of the generalized  form  
$
\mathbf{V }\o(r_j)  = -   i\mathbf{z}\o_j  
\mathbf{U}\o(r_j)$ $ (j=1,2)$ 
where $\mathbf{z}\o_j$ $(j=1,2)$ are known quantities.  
The conditional impedance  $ \mathbf{z}\o (r)$ is determined (numerically) by integrating \eqref{132} from (say) $r=r_1$ with initial condition $ \mathbf{z}\o (r_1)={\mathbf z}\o_1$
to give
 \beq{-42}
\mathbf{z}\o (r)= i\big( \mathbf{M}_{3}( r ,r_1)-i\mathbf{M}_{4}( r ,r_1){\mathbf z}\o_1 \big) \big( \mathbf{M}_{1}( r ,r_1)-i\mathbf{M}_{2}( r ,r_1){\mathbf z}\o_1 \big)^{-1} .
\eeq 
The interface condition at $r=r_2$ requires that 
\beq{-466}
-   \mathbf{z}\o_2  
\mathbf{U}\o(r_2) = - \mathbf{z}\o (r_2)\mathbf{U}\o(r_2),
\nonumber
\eeq
which implies the dispersion equation   
\beq{-467}
\det\big\{ i\big( \mathbf{M}_{3}( r_2 ,r_1)-i\mathbf{M}_{4}( r_2 ,r_1){\mathbf z}\o_1 \big) \big( \mathbf{M}_{1}( r_2 ,r_1)-i\mathbf{M}_{2}( r_2 ,r_1){\mathbf z}\o_1 \big)^{-1}
-   \mathbf{z}\o_2 \big\}=0.
\eeq
Variants on this equation may be obtained using the two-point impedance instead of the matricant.  Thus, from 
eq.\ \eqref{012} 
 we have the equivalent condition  
\beq{-404}
 \det\big\{ 
 \mathbf{z}\o_2 + \mathbf{Z}_{4}( r_2 ,r_1) + \mathbf{Z}_{3}( r_2 ,r_1)
\big(  {\mathbf z}\o_1-\mathbf{Z}_{1}( r_2 ,r_1)  \big)^{-1}\mathbf{Z}_{2}( r_2 ,r_1)
 \big\}=0.
 \nonumber
\eeq

The examples considered in this section illustrate the usefulness of wave impedance matrices for cylinders of finite and infinite radial extent.   Solutions to  problems of practical concern can be formulated concisely in terms of impedance matrices, such as the dispersion equation for guided waves, and the scattering of waves from a cylindrical region.   Calculation of the impedance matrices is relatively straightforward 
using the matricant or two-point impedance matrices (see \cite{Shuvalov03}), but only as 
long as   the points $r=0$ or $r=\infty$ are not involved; otherwise the solid cylinder impedance and/or radiation impedance matrices are required. 
 Determination  of the solid-cylinder impedance matrix $ \mathbf{Z}\o(r)$ is   discussed  next.

\section{The solid impedance matrix  }\label{sec4}

In this section we develop methods to calculate the solid-cylinder impedance matrix for a
radially inhomogeneous  cylindrically anisotropic cylinder with material at  $r=0$.   Two principle approaches are considered: a semi-explicit solution as a Frobenius series, and an implicit solution in terms of a differential Riccati  equation. 

Unlike the conditional impedance which can be determined directly from the matricant  $\mathbf{M}$ along with the prescribed reference value,  the matricant is not of direct use here because of its  divergence   at $r=0 $.  This introduces the need to  identify  `physical' and `nonphysical' constituents of  the solution near $r=0$, which is performed explicitly for the Frobenius solution.  In the Riccati approach the displacement and traction fields are not considered explicitly and the divergence  at $r=0$ is taken care of by  the initial value of the impedance. 
The Frobenius solution is considered first. 

\subsection{Frobenius expansion }

We take advantage of the fact that the fundamental 
solution  can formally be written  in terms of
a  Frobenius series, which is  an explicit  one-point solution valid  at any $r$ (including $r=0$). As a result, the Frobenius-series approach  provides a constructive 
 definition of ${\mathbf Z}\o (r)$.    
  The Frobenius series solution can be obtained   via a recursive
procedure with the number of numerically  required terms increasing with $r$.    Before we present the formal solution for ${\mathbf Z}\o (r)$  we review and develop  some properties of the Frobenius series for cylindrically anisotropic materials, following  the   analysis  in \cite{Shuvalov03a}.

\subsubsection{Background material} \label{sec4.1.1}

 The Frobenius solution   is based on 
 the    
 integral matrix solution $\pmb{\mathcal{N}}%
\left( y\right) =\left\Vert \pmb{\eta }_{1},...,\pmb{\eta }_{6}\right\Vert $ of Eq. \eqref{130}, which   can always be defined through the Frobenius series  for any $r\ge 0$. 
The pivotal role in constructing this series  belongs to the eigenspectrum of
the $6\times 6$ matrix $\mathbf{g}_{0}\left( 0\right) $ with the symmetry%
\begin{equation}
\mathbf{g}_{0}=-\mathbf{Tg}_{0}^{+}\mathbf{T},  \label{1}
\end{equation}
which follows from \eqref{056}. Denote the eigenvalues and eigenvectors of $%
\mathbf{g}_{0}\left( 0\right) $ by $\lambda _{\alpha }$ and 
$\pmb{\gamma}_{0\alpha }=\left( \mathbf{a}_{\alpha },\mathbf{l}_{\alpha }\right) ^{%
\mathrm{T}}$ ($\alpha =1,...,6$), and introduce the matrix $\mathbf{\Gamma }%
_{0}=\left\Vert \pmb{\gamma }_{01},...,\pmb{\gamma }_{06}\right\Vert $.
Barring extraordinary exceptions, if $n>1$ then 
 (i)  no two  eigenvalues $\lambda _{\alpha }$ of $\mathbf{%
g}_{0}\left( 0\right) $ differ by an integer, and (ii) all  $\lambda _{\alpha }$  normally are distinct (and nonzero).  Let us  first consider this case $n>1$, otherwise see \S\ref{4.1.2}. By virtue of  (i), the 
 integral matrix  may be formulated as   
\begin{equation}
\pmb{\mathcal{N}}\left( r\right) =\mathbf{D}\left( r\right) \mathbf{%
\mathbf{\Gamma }}_{0}\mathbf{C} ,\quad 
\mathbf{D}\left( r\right) =\mathbf{I} + \sum\limits_{m=1}^{\infty }\mathbf{D}%
_{m}r^{m}  ,  \label{882}
\end{equation}%
where $\mathbf{C}$ 
is the Jordan form 
of the matrix $r^{\mathbf{g}_{0}\left( 0\right) }$ which is diagonal when  (ii) holds, and $\mathbf{D}\left( r\right) $
is
defined recursively through $\mathbf{G}\left( r\right) $ \cite[Eqs. (9)-(13)]{Shuvalov03a}. 

The arguments underlying eq. \eqref{91} imply
that the matrix $\pmb{\mathcal{N}}^{+}\left( r\right) \mathbf{T}\pmb{\mathcal{N}}\left(
r\right) $    is a constant independent of $r$, and according to    eq.\ \eqref{010}   
this matrix defines the flux properties of the  constituents $\pmb{\eta }_1$, $\ldots$ , 
$\pmb{\eta }_6$, 
of $\pmb{\mathcal{N}}\left(
r\right) $.  For the present purposes we wish to split them  
into a pair of triplets: a physical set $(\alpha =1,2,3)$ and a nonphysical   triplet 
$(\alpha =4,5,6)$, where the only non-zero flux interactions occur between $\alpha$ and $\alpha +3$   $(\alpha =1,2,3)$, thus  ensuring the crucial property that $\pmb{\mathcal{N}}^{+}\left( r\right) \mathbf{T}\pmb{\mathcal{N}}\left(
r\right) $ has  
  nonzero elements  confined to the main diagonal of the off-diagonal blocks.
 The partitioning is accomplished  
through appropriate arrangement   of 
the eigenspectrum of $\mathbf{g}_{0}\left( 0\right) $ as 
\begin{equation}
\lambda _{\alpha }=-\lambda _{\alpha +3}^{\ast },\quad \Re\lambda _{\alpha
}>0,\quad \alpha =1,2,3,  \label{883}
\end{equation}%
\cite[Eq.  (44)]{Shuvalov03a}. 
Combining    Eqs.  (\ref{1}) and (\ref{883}) and adopting the  normalization 
$\pmb{\gamma}_{0\alpha }^+ {\mathbf T}\pmb{\gamma}_{0\alpha +3}=1$ 
yields the orthogonality/completeness relation for the eigenvectors in the form
\begin{equation}
\mathbf{\Gamma }_{0}^{+}\mathbf{T\mathbf{\Gamma }}_{0}=\mathbf{T} . \label{884}
\end{equation}%
It follows from Eqs. (\ref{1}) through (\ref{884}) that $\pmb{\mathcal{N}}^{+}\left( 0\right) \mathbf{T}\pmb{\mathcal{N}}\left( 0\right) =\mathbf{T}$ and hence the  flux matrix at $r$ is $\mathbf{T}$, 
\begin{equation}
\pmb{\mathcal{N}}^{+}\left( r\right) \mathbf{T}\pmb{\mathcal{N}}\left( r\right)
=\mathbf{T}  
\quad
\big(\Rightarrow \, 
\pmb{\mathcal{N}}\left( r\right) \mathbf{T}\pmb{\mathcal{N}}^{+}\left( r\right)
=\mathbf{T}
\big)
.\label{885}
\end{equation}  
Note that \eqref{884} yields $\mathbf{D}^{+}\mathbf{TD}=\mathbf{T} $. 


In order to further clarify the structure of $\pmb{\mathcal{N}}$  we represent the 6$\times $6 matrices 
$\mathbf{D}$, $\mathbf{\Gamma }_{0}$ and $\mathbf{C} $ 
in terms of $3\times 3$ 
 submatrices, 
\beq{-332}
\mathbf{D}\left( r\right) =
\begin{pmatrix}
\mathbf{D}_{1} & \mathbf{D}_{2} \\  & \\
\mathbf{D}_{3} & \mathbf{D}_{4}%
\end{pmatrix}%
  ,\ \mathbf{\Gamma }_{0}=
  \begin{pmatrix}
\mathbf{A}_{1} & \mathbf{A}_{2} \\ & \\
\mathbf{L}_{1} & \mathbf{L}_{2}%
\end{pmatrix}
 ,\ \mathbf{C} = 
\begin{pmatrix}
\mathrm{diag}\left( r^{\lambda _{\alpha }}\right)  & {\mathbf{0}} \\ & \\
{\mathbf{0}} & \mathrm{diag}\left( r^{-\lambda _{\alpha }^{\ast
}}\right) 
\end{pmatrix}, 
\eeq  
where $\alpha =1,2,3$ and $\mathbf{C}$ is diagonal for $n>1$. 
The integral matrix consequently  has  block structure
\beq{6}
\pmb{\mathcal{N}}\left( r\right) =
 \begin{pmatrix}
\widehat{\mathbf{U}}_{1} & \widehat{\mathbf{U}}_{2} \\ & \\
\widehat{\mathbf{V}}_{1} & \widehat{\mathbf{V}}_{2}%
 \end{pmatrix}= \begin{pmatrix}
\mathbf{D}_{1} & \mathbf{D}_{2} \\ & \\
\mathbf{D}_{3} & \mathbf{D}_{4}%
 \end{pmatrix} \begin{pmatrix}
\mathbf{A}_{1}\mathrm{diag}\left( r^{\lambda _{\alpha }}\right)  & \mathbf{A}%
_{2}\mathrm{diag}\left( r^{-\lambda _{\alpha }^{\ast }}\right)  \\ & \\
\mathbf{L}_{1}\mathrm{diag}\left( r^{\lambda _{\alpha }}\right)  & \mathbf{L}%
_{2}\mathrm{diag}\left( r^{-\lambda _{\alpha }^{\ast }}\right) 
 \end{pmatrix}.
 \eeq
Note in particular that   the integral matrix $\pmb{\mathcal{N}}\left( r\right) $
consists of two distinct 6$\times $3 matrices, 
\begin{equation}
\begin{split} &
\begin{pmatrix}
\widehat{\mathbf{U}}_{1}\left( r\right)  \\ 
\widehat{\mathbf{V}}_{1}\left( r\right) 
\end{pmatrix} 
=\left\Vert \pmb{\eta }_{1},\pmb{\eta }_{2},\pmb{\eta }%
_{3}\right\Vert =\mathbf{D}\left\Vert \pmb{\gamma }_{01},\pmb{\gamma }%
_{02},\pmb{\gamma }_{03}\right\Vert \mathrm{diag}\left( r^{\lambda
_{\alpha }}\right) ,
\\ &
\begin{pmatrix}
\widehat{\mathbf{U}}_{2}\left( r\right)  \\ 
\widehat{\mathbf{V}}_{2}\left( r\right) 
\end{pmatrix}
=\left\Vert \pmb{\eta }_{4},\pmb{\eta }_{5},\pmb{\eta }%
_{6}\right\Vert =\mathbf{D}\left\Vert \pmb{\gamma }_{04},\pmb{\gamma }%
_{05},\pmb{\gamma }_{06}\right\Vert \mathrm{diag}\left( r^{-\lambda
_{\alpha }^{\ast }}\right) ,
\end{split}
\label{888}
\end{equation}%
the former with the columns $\pmb{\eta }_{\alpha }\left( r\right) $
tending to zero at $r\rightarrow 0,$ and the latter with columns $\mathbf{%
\eta }_{\alpha +3}\left( r\right) $ diverging at $r\rightarrow 0$. 
The  block structure of eqs.\ (\ref{884}) and (\ref{885}) is  
\bal{887}
\mathbf{\Gamma }_{0}^{+}\mathbf{T\mathbf{\Gamma }}_{0} &=
 \begin{pmatrix}
\mathbf{A}_{1}^+ & \mathbf{L}_{1}^+ \\ & \\
\mathbf{A}_{2}^+ & \mathbf{L}_{2}^+%
\end{pmatrix}
 \begin{pmatrix}
\mathbf{L}_{1} & \mathbf{L}_{2} \\ & \\
\mathbf{A}_{1} & \mathbf{A}_{2}%
\end{pmatrix}  = 
\begin{pmatrix}
{\mathbf{0}} & \mathbf{I} \\ & \\
\mathbf{I} & {\mathbf{0}}%
\end{pmatrix},
 \nonumber 
 \\ & \\
\mathbf{\pmb{\mathcal{N}}^{+}}\left( r\right) \mathbf{T}\pmb{\mathcal{N}}\left(
r\right) &= 
\begin{pmatrix}
\widehat{\mathbf{U}}_{1}^{+} & \widehat{\mathbf{V}}_{1}^{+} \\ & \\ 
\widehat{\mathbf{U}}_{2}^{+} & \widehat{\mathbf{V}}_{2}^{+}%
\end{pmatrix}  
\begin{pmatrix}
\widehat{\mathbf{V}}_{1} & \widehat{\mathbf{V}}_{2} \\ & \\
\widehat{\mathbf{U}}_{1} & \widehat{\mathbf{U}}_{2}%
\end{pmatrix}  = 
\begin{pmatrix}
{\mathbf{0}} & \mathbf{I} \\ & \\
\mathbf{I} & {\mathbf{0}}%
\end{pmatrix} . \nonumber 
\end{align}%
The latter 
explicitly  shows that 
the normal energy flux of 
the displacement-traction wave field $\pmb{\eta }\left( r\right) $
comprising an arbitrary superposition of either the three modes $\mathbf{%
\eta }_{\alpha }\left( r\right) $ or three modes $\pmb{\eta }_{\alpha
+3}\left( r\right) $  with $\alpha =1,2,3$ is zero at any $r$. 
This specific arrangement of $\pmb{\mathcal{N}}$ may be interpreted as the generalization of the isotropic case with solutions  cast in terms of the cylinder functions $J_n$ and $-iY_n$, corresponding to the physical and nonphysical triplets respectively, each of which yields zero flux individually.   
This partitioning  will be crucial in developing  an  explicit solution for the solid impedance matrix. 

\subsubsection{Overview of the cases $n=0$ and $n=1$}\label{4.1.2}

Let us return to the two  assumptions made above which are that (i) no two
eigenvalues $\lambda _{\alpha }\o$ of $\mathbf{g}%
_{0}\o\left( 0\right) $ differ by an integer and (ii) all $%
\lambda _{\alpha }\o$ are distinct, hence $\mathbf{g}%
_{0}\o\left( 0\right) $ is semisimple (diagonalizable). Violating (i) invalidates the relatively simple form (\ref{882}) of
the Frobenius fundamental solution to the governing equation, 
see \cite{Coddington}. Violation of (ii), or more precisely,  the occurrence of degenerate $\lambda
_{\alpha }\o$   that makes $\mathbf{g}_{0} \left( 0\right) $ non-semisimple,  alters the orthogonality/completeness relations and the composition of $\pmb{\mathcal{N}}$ given above for $n>1$.  
 The cases affected are $n=0$ (axisymmetric
modes) and $n=1$ (lowest-order flexural modes): specifically, the property (i) does not
hold for $n=0,$ and the property (ii) does not hold for both $n=0$ and $n=1$.
From a physical point of view, the cases $n=0,~1$ stand out because they are related to the
rigid-body motions producing zero stresses \cite[Eq. (19)]{Shuvalov03a}. Note also
that $\mathbf{g}_{0}\o\left( 0\right) $ admits a zero
eigenvalue iff $n=0$, $1$ 
\cite[Eq. (30)$_3$]{Shuvalov03a}
and that $\lambda ^{\left(0,1\right) }=0$ is always a double eigenvalue   rendering $\mathbf{g}
_{0}^{\left( 0,1\right) }\left( 0\right) $ non-semisimple. 

Consider the axisymmetric case $n=0.$ The six eigenvalues $\lambda _{\alpha }^{(0)}$ of  $\mathbf{g}_{0}^{(0)} \left(
0\right) $ are $\lambda _{\alpha }^{(0)} =\left\{ 0,0,\pm 1,\pm
\kappa \right\} ,$ where $\kappa =1$ for   trigonal or tetragonal symmetry with $%
c_{16}=0$  \cite[Eqs. (3.12), (3.13)]{Ting96a}. 
It is seen that, whatever the symmetry, the
set of $\lambda _{\alpha }^{\left( 0\right) }$ includes pairs different by
an integer. As a result, the integral matrix $\pmb{\mathcal{N}}\left( r\right) $ is now defined through  $\mathbf{g}_{0}^{(0)} \left(0\right) $  in a rather intricate form elucidated in   \cite[Eqs. (A2), (A.4)]{Shuvalov03a}. 
This observation is essential for treating inhomogeneous
and low-symmetry homogeneous cylinders.
At the same time,  if the cylinder is homogeneous and has   orthorhombic
or higher symmetry with the exception of   trigonal and tetragonal with $c_{16}=0$,
then $\pmb{\mathcal{N}}\left( r\right) $ decouples
into the solutions described by Bessel functions and/or by a    simple 
Frobenius form (\ref{882})\footnote{%
Orthorhombic or higher symmetry enables uncoupling of the pair of
torsional modes described by the Bessel solutions stemming from $\lambda
^{\left( 0\right) }=\pm 1.$ The four sagittal modes are associated with $%
\lambda ^{\left( 0\right) }=\left\{ 0,0,\pm \kappa \right\} ,$ where $\kappa
\neq 1$ for   symmetry lower than the trigonal or tetragonal with $c_{16}=0
$. When $\kappa =1,$ so that the above quartet of $\lambda ^{\left( 0\right)
}$ involves pairs with an integer difference, the sagittal problem admits
explicit Bessel solutions for the isotropic or transverse isotropic symmetry
due to uncoupling of potentials. Note that double eigenvalues $\lambda ^{\left( 0\right) }=\pm 1$
at $\kappa =1$ do not bring   non-diagonal blocks into the Jordan form of 
$\mathbf{g}_{0}^{\left( 0\right) }\left( 0\right) .$}.

Consider the case $n=1.$ The  matrix $\mathbf{g}_{0}^{\left( 1\right)
}\left( 0\right) $ has a doubly degenerate eigenvalue $\lambda ^{\left(
1\right) }=0$ which makes $\mathbf{g}_{0}^{\left( 1\right) }$ non-semisimple
\cite[Eq. (36)]{Shuvalov03a}. This does not preclude taking $\pmb{\mathcal{N}%
} \left( r\right) $ in the form (\ref{882}) but the matrix $\mathbf{C}$
is now not diagonal. 
As a
result, the triplet $\alpha =1,2,3$ of physical modes (with one of the modes 
$\pmb{\eta }_{\alpha }^{\left( 1\right) }$ associated with $\lambda
^{\left( 1\right) }=0$) retains its form \eqref{888}$_{1}$, whereas the
nonphysical triplet $\alpha =4,5,6$ is no longer of the form \eqref{888}$_{2}$ due
to one of its modes involving both eigenvectors, the proper and the
generalized ones $\pmb{\gamma}^{(1)}$ and $\pmb{\widetilde\gamma}^{(1)}$, associated with $\lambda ^{\left( 1\right) }=0$ 
\cite[Eqs. (51), (61)]{Shuvalov03a}. It is thus evident that the physical modes satisfy
the same orthogonality/completeness relations as for $n>1$; moreover, subject to the optional condition 
$\pmb{\widetilde\gamma}^{(1)+} \mathbf{T} \pmb{\gamma}^{(1)}=0$, the
nonphysical modes may be shown to do so as well. The relations \eqref{884} and \eqref{885}  for the case $n=1$ are accordingly modified into a slightly different
form%
\begin{equation}
\mathbf{\Gamma }_{0}^{+}\mathbf{T\Gamma }_{0}=\mathbf{E},
\qquad  
{\pmb{\mathcal{N}}^{+}}\left( r\right) \pmb{T\mathcal{N}}\left( r\right) =\mathbf{E}
\qquad %
\left( n=1\right) ,  \label{7.0}
\end{equation}%
which differs from \eqref{884} and \eqref{885} only in the replacement of the
right-hand matrix $\mathbf{T}$ by $\mathbf{E}$, whose nonzero elements are
also confined to the main diagonal of the off-diagonal blocks but they
cannot now be all normalized to 1 \cite[Eq. (49)]{Shuvalov03a}.

The overall conclusion is that both cases $n=0$ and $n=1$ preserve the  
partitioning  of the six linear independent Frobenius solutions $\pmb{%
\eta }_{\alpha }\left( r\right) =\left( \mathbf{U}%
_{\alpha },\mathbf{V}_{\alpha }\right) ^{\mathrm{T}}$ within $\pmb{%
\mathcal{N}}\left( r\right) $ ($\alpha =1,...,6$) into the physical and
nonphysical triplets $\alpha =1,2,3$ and $\alpha =4,5,6.$ The partitioning
is based on  \eqref{883} supplemented by including the (double) eigenvalue $%
\lambda ^{\left( 0,1\right) }=0$. 
The vectors $\mathbf{U}_{\alpha }\left(
r\right) $ and $\mathbf{V}_{\alpha }\left( r\right) $ are certainly regular
at $r\rightarrow 0$ for both $n=0$, $1$, although the limiting trend for $n=0$ is
not of the form that results from \eqref{882}, see \cite[Eq. (A4)]{Shuvalov03a}. Equations \eqref{888} and
 \eqref{887},  which are valid for any $n>0$, enable treating the solid
cylinder impedance $\mathbf{Z}\left( r\right) $ for $n=1$ on the same
grounds as for the `ordinary' case $n>1$.  The impedance $\mathbf{Z}%
\left( r\right) $ for $n=0$ needs special attention because the case $n=0$
may not satisfy \eqref{882}. 
  We are now ready to derive the explicit form of the solid-cylinder impedance for all $n$.

\subsection{Explicit solution of the solid-cylinder impedance}

\subsubsection{The solid-cylinder impedance for arbitrary $n$}

The definition \eqref{21} of the solid-cylinder impedance $\mathbf{Z}\o\left(
r\right)$ tacitly assumes   ${\mathbf U}\o(r)$ and ${\mathbf V}\o(r)$ are regular function of $r$.  
This is always so for $%
\pmb{\eta } \left( r\right) =\left( \mathbf{U},\mathbf{V%
}\right) ^{\mathrm{T}}$ comprising an arbitrary superposition of,
specifically, the physical Frobenius modes $\pmb{\eta }_{\alpha
}\left( r\right) =\left( \mathbf{U}_{\alpha },\mathbf{V}%
_{\alpha }\right) ^{\mathrm{T}}$ which satisfy eq.\ \eqref{883} supplemented by the
option $\lambda ^{\left( 0,1\right) }=0$ for $n=0$, $1$ (see \S \ref{4.1.2}). Thus
the solid-cylinder impedance may be defined by any  of the  equivalent
expressions%
\begin{equation}
\mathbf{V}_{\alpha }\left( r\right) =-i\mathbf{Z}\o\left( r\right) \mathbf{U}%
_{\alpha }\left( r\right) \ \ \ \left( \alpha =1,2,3\right)\, \,  \Leftrightarrow\, \, 
\ \mathbf{V}\left( r\right) =-i\mathbf{Z}\o\left( r\right) \mathbf{U}\left(
r\right) 
\, \,  \Leftrightarrow\, \, 
 \mathbf{Z}\o \left( r\right) =i\widehat{\mathbf{V}}_{1}\left( r\right) \widehat{\mathbf{U}}_{1}^{-1}\left( r\right).  \label{901}
\end{equation}%
This yields a finite value if $\det \widehat{\mathbf{U}}_{1}\left( r\right)
\neq 0$,  otherwise the impedance is associated with a `rigid' condition at $r$ (conversely, the determinant of its inverse  - the admittance matrix - is zero).  The occurrence of infinities is in no way  anomalous but rather a natural consequence  of the definition of the impedance matrix.
 
 Consider first $n>0$.  
Based on the definition \eqref{901} and the representation \eqref{888}$_1$ for the $3\times 3$ matrices 
$\widehat{\mathbf{U}}_{1}$ and $\widehat{\mathbf{V}}_{1}$, we obtain an  alternative form  for the solid-cylinder impedance,
\beq{092}
  \mathbf{Z}\o \left( r\right) =i \big( {\mathbf D}_3(r)-i{\mathbf D}_4(r) {\mathbf Z}_0\big)
\big( {\mathbf D}_1(r) -i {\mathbf D}_2(r)  {\mathbf Z}_0 \big)^{-1}
\quad\text{where  } {\mathbf Z}_0   = i  {\mathbf L}_1{\mathbf A}_1^{-1}.
\eeq
Hermiticity of the solid-cylinder impedance follows from \eqref{887}$_2$ and \eqref{7.0}$_2$ which imply 
that  $\widehat{\mathbf{U}}_{1}^{+}\widehat{\mathbf{V}}_{1}+\widehat{\mathbf{V%
}}_{1}^{+}\widehat{\mathbf{U}}_{1}=-i\widehat{\mathbf{U}}_{1}^{+}\left( 
\mathbf{Z-Z}^{+}\right) \widehat{\mathbf{U}}_{1}={\mathbf{0}}$, whence%
\begin{equation}
\mathbf{Z}\o\left( r\right) ={\mathbf{Z}\o}^{+}\left( r\right) .
\label{12}
\end{equation}
The expression \eqref{092} is reminiscent of the representation of the conditional impedance, e.g. eq.\ \eqref{-42}, except that the role of the two-point matricant ${\mathbf M}(r,r_0)$ is replaced by
${\mathbf D}(r)$.  
Note that $\det \mathbf{A}_{1}\neq 0$ may be deduced  
 from the integral representation of ${\mathbf Z}_0$, see eq. \eqref{7.22}$_1$, by reasoning similar to \cite{Barnett85}: that if two 
 of  the eigenvectors are parallel, say $\mathbf{a}_{1 }$ and $\mathbf{a}_{2 }$, then so are the 
 traction vectors $\mathbf{l}_{\alpha }= -i \mathbf{Z}_0\mathbf{a}_{\alpha }$ counter to 
 the assumed linear independence of $\pmb{\gamma}_{1 }$ and $\pmb{\gamma}_{2 }$.

Now consider   $n=0$.  
Violation of eq. (\ref{882}) for $\mathbf{\mathcal{N}}%
^{\left( 0\right) }\left( r\right) $ invalidates the definition \eqref{092}$_2$
for the central impedance $\mathbf{Z}_{0}^{\left( 0\right) }.$ At the same
time, $\mathbf{Z}_{0}^{\left( 0\right) }$ can readily be found by means of a
direct derivation given in \S\ref{5.3.2}, specifically eq. \eqref{-47}, which is    
clearly 
Hermitian regardless of anisotropy.  Consequently  $\mathbf{Z}^{\left(
0\right) }\left( r\right) $ is Hermitian for any $r$ due to 
the self-adjoint property of the  differential Riccati equation of which $\mathbf{Z}^{\left(
0\right) }\left( r\right) $ is the unique physical solution (see \S\ref{4.3}).

\subsubsection{The link between the solid-cylinder and the conditional impedances }\label{4.2.2}

 It should be  evident from the previous discussion that 
 $\mathbf{Z}\o\left( r\right) $ can formally be
defined as the conditional
impedance $\mathbf{z}\o\left( r\right) $ with  initial value at $%
r_{0}\rightarrow 0$.  Assume for brevity that $n>1$, then using   the representation 
$\mathbf{M} ( y,y_0 ) = \pmb{\mathcal{N}}(y) \pmb{\mathcal{N}}^{-1}(y_0) $ for the matricant 
and eqs.\  (\ref{885}), \eqref{6},  we have 
\begin{equation}
\mathbf{M}\left( r,r_{0}\right) \underset{r_{0}\rightarrow 0}{\rightarrow }%
\begin{pmatrix}
\widehat{\mathbf{U}}_{1}(r) & \widehat{\mathbf{U}}_{2}\left(
r\right)  \\ 
\widehat{\mathbf{V}}_{1}(r) & \widehat{\mathbf{V}}_{2}\left(
r\right) 
\end{pmatrix} 
\begin{pmatrix}
\widehat{\mathbf{V}}_{2}^{+}\left( r_{0}\right)  & \widehat{\mathbf{U}}%
_{2}^{+}\left( r_{0}\right)  \\ 
{\mathbf{0}} & {\mathbf{0}}%
\end{pmatrix}
=\begin{pmatrix}
\widehat{\mathbf{U}}_{1}(r)\widehat{\mathbf{V}}_{2}^{+}\left(
r_{0}\right)  & \widehat{\mathbf{U}}_{1}(r)\widehat{\mathbf{U}}%
_{2}^{+}\left( r_{0}\right)  \\ 
\widehat{\mathbf{V}}_{1}(r)\widehat{\mathbf{V}}_{2}^{+}\left(
r_{0}\right)  & \widehat{\mathbf{V}}_{1}(r)\widehat{\mathbf{U}}%
_{2}^{+}\left( r_{0}\right) 
\end{pmatrix}_{r_{0}\rightarrow 0}.  \label{815}
\nonumber
\end{equation}%
This illustrates that even though  the matricant $\mathbf{M}\left( r,r_{0}\right) $  diverges
at $r_{0}\rightarrow 0$, as expected, it provides the correct limit 
\begin{equation}
\mathbf{z}\o\left( r\right) =i\left( \mathbf{M}_{3}-i\mathbf{M}_{4}\mathbf{z}\o%
\left( r_{0}\right) \right) \left( \mathbf{M}_{1}-i\mathbf{M}_{2}\mathbf{z}\o%
\left( r_{0}\right) \right) ^{-1}\underset{r_{0}\rightarrow 0}{\rightarrow }i%
\widehat{\mathbf{V}}_{1}\left( r\right) \widehat{\mathbf{U}}_{1}^{-1}\left(
r\right) =\mathbf{Z}\o\left( r\right) .  \label{816}
\end{equation}
Formal consistency requires     the limiting value of $%
\mathbf{z}\o\left( r_{0}\right) $ at $r_{0}\rightarrow 0$  be set  equal to $\mathbf{%
Z}\o_{0}$.   However, the definition (\ref{816}) of $\mathbf{Z}\o\left( r\right) $
is actually of no value for practical calculations because of the divergence
of $\mathbf{M}\left( r,r_{0}\right) $ at $r_{0}\rightarrow 0$.

At the same time,   in the limit as $r\rightarrow 0$   the conditional
impedance $\mathbf{z}\o\left( r\right) $   with any initial value $%
\mathbf{z}\o\left( r_0\right) $ such that $|\mathbf{z}\o( r_{0})
-\mathbf{Z}\o( r_{0})|>0$, should tend to the 
nonphysical 
central impedance $\mathbf{Z}_{\mathrm{np}}\left( 0\right) =i\mathbf{L}_{2}%
\mathbf{A}_{2}^{-1}$.   Similarly to (\ref{815}), 
\begin{equation}
\mathbf{M}\left( r,r_{0}\right) \underset{r\rightarrow 0}{\rightarrow }%
\begin{pmatrix}
{\mathbf{0}} & \widehat{\mathbf{U}}_{2}\left( r\right)  \\ 
{\mathbf{0}} & \widehat{\mathbf{V}}_{2}\left( r\right) 
\end{pmatrix} _{r\rightarrow 0}
\begin{pmatrix}
\widehat{\mathbf{V}}_{2}^{+}\left( r_{0}\right)  & \widehat{\mathbf{U}}%
_{2}^{+}\left( r_{0}\right)  \\ 
\widehat{\mathbf{V}}_{1}^{+}\left( r_{0}\right)  & \widehat{\mathbf{U}}%
_{1}^{+}\left( r_{0}\right) 
\end{pmatrix}
=\begin{pmatrix}
\widehat{\mathbf{U}}_{2}\left( r\right) \widehat{\mathbf{V}}_{1}^{+}\left(
r_{0}\right)  & \widehat{\mathbf{U}}_{2}\left( r\right) \widehat{\mathbf{U}}%
_{1}^{+}\left( r_{0}\right)  \\ 
\widehat{\mathbf{V}}_{2}\left( r\right) \widehat{\mathbf{V}}_{1}^{+}\left(
r_{0}\right)  & \widehat{\mathbf{V}}_{2}\left( r\right) \widehat{\mathbf{U}}%
_{1}^{+}\left( r_{0}\right) 
\end{pmatrix} _{r\rightarrow 0}.  \label{13}
\nonumber
\end{equation}%
Hence, from     \eqref{-42}, 
\beq{814}
\mathbf{z}\o\left( r\right) =i\left( \mathbf{M}_{3}-i\mathbf{M}_{4}\mathbf{z}\o%
\left( r_{0}\right) \right) \left( \mathbf{M}_{1}-i\mathbf{M}_{2}\mathbf{z}\o%
\left( r_{0}\right) \right)^{-1}
\underset{r\rightarrow 0}{\rightarrow }i%
\left[ \widehat{\mathbf{V}}_{2}\left( r\right) \widehat{\mathbf{U}}%
_{2}^{-1}\left( r\right) \right] _{r\rightarrow 0}=i\mathbf{L}_{2}\mathbf{A}%
_{2}^{-1}. 
\eeq
  If $\mathbf{z}\o( r_{0}) $ is  precisely   the solid impedance at $r_0$ 
then   \eqref{814}$_1$ 
 reproduces  the solid impedance,  $\mathbf{z}\o\left( r\right) = \mathbf{Z}\o\left( r\right) $ for $r>0$.   
But  the limit at $r=0$,   formally $\mathbf{Z}\o\left( 0\right)
=\mathbf{Z}_0$, cannot be achieved  in practice, a reflection of the fact that the matricant based solution \eqref{023} in cylindrical coordinates  is uniquely ill-posed  at this point (see also \S\ref{sec4.4}).

\subsection{Riccati equation solution}\label{4.3}

An alternative to the Frobenius approach  is to consider   ${\mathbf Z}(r)$  as a   solution of the differential Riccati equation with 
initial value ${\mathbf Z_0}$ extended to the case when the initial value occurs at $r=0$.  The solid-cylinder impedance is then the  solution of the initial value problem,    
\beq{228}
 \begin{split}
 r\frac{\dd {\mathbf Z}\o}{\dd r}=
 \big[  {\mathbf Z}\o +\big(\widetilde{\mathbf R} + ik_z r {\mathbf P}\big)^+\big]
 {\mathbf{\widehat{Q}}}^{-1} \big[  {\mathbf Z}\o + \widetilde{\mathbf R} + ik_z r {\mathbf P} \big]
 - {\mathbf B}(r)  , \quad r \ge 0; 
 \quad 
 \mathbf{Z}\o ( 0)
=  \mathbf{Z}\o_0,
 \end{split}
\eeq
where ${\mathbf B}(r) $ is defined in   eq.\  \eqref{224}. 
The central-impedance matrix  ${\mathbf Z_0}$, as discussed in the previous subsection,   is defined by the eigenvectors of ${\mathbf g}_0(0)$, see \eqref{092}$_2$.   Alternatively, noting that a nonphysical singularity is introduced unless the right hand side of \eqref{228}$_1$  vanishes at $r=0$, we conclude that the central impedance must   satisfy the algebraic Riccati equation 
\beq{25}
  \big(  {\mathbf Z}\o_0 + \widetilde{\mathbf R}_0^+ \big)
 {\mathbf{\widehat{Q}}}_0^{-1} \big(  {\mathbf Z}\o_0 + \widetilde{\mathbf R}_0   \big) 
  - \widetilde{\mathbf T}_0 ={\mathbf{0}} .
\eeq

 While it is expected that the solution $ {\mathbf Z}\o(r)$ is well behaved in some finite neighbourhood of $r=0$, the Riccati solution will inevitably develop singularities.  These are associated with 
guided waves of a    cylinder of radius $r$ with clamped surface (zero displacement condition).   For  given $\omega$ and $k_z$, the singularities  occur at  values of $r$ such that $\det \widehat{\mathbf{U}}_{1}\left( r\right)
= 0$ (see eq.\ \eqref{901}).  Thus,  
 one can integrate the differential Riccati equation     only as far as the first singularity at (say) $r=r_*$.  The problem is evident from the example of the out-of-plane  impedance  derived  in \eqref{33}$_2$, 
 $Z\o_z(r,0) =     -c_{44} \, k_2 r   {J_n'(k_2 r)}/{J_n(k_2 r)} $, 
  which blows up when $k_2 r$ is a zero of the Bessel function $J_n$.    
  The effect of singularities may be   circumvented in practice by  integrating the Riccati equation  to some     finite $r$   short of the first singularity and then to switch to some other solution method that is regular in the vicinity of $r=r_*$.  One approach   \cite{Biryukov85}  is to consider the admittance (inverse of impedance) $\mathbf{Y}(r) = \mathbf{Z}^{-1}(r)$ which will be well behaved at $r=r_*$. Its differential Riccati equation, which is easily found from \eqref{228}, can therefore be integrated without incident through the singularity at $r=r_*$, but the admittance then has its own singularities  at positions different from those of the impedance,  so in general this approach requires switching back and forth  between two  Riccati equations.   While  certainly feasible, the procedure is complicated by the fact that one does not know the singularities {\it a priori}.  
Note that the admittance Riccati equation is not suitable for starting at $r=0$ because as discussed in the next section $\det \mathbf{Z}_0 =0$ and hence $\mathbf{Y}_0 = \mathbf{Y}(0)$ is undefined for $n=0$, $1$.  

 A more practical approach  to deal with the unavoidable singularity problem   is to  
      use the Riccati solution  to generate  initial conditions for the full $6\times 6$ system at $r=r_1 <r_*$, with which one can  integrate (again numerically) to arbitrary $r> r_1$ using \eqref{-42}.
In practice one only needs to solve for a $6\times 3$ matrix ${\mathbf \Omega}(r) $, satisfying 
\beq{496}
 \frac{\dd}{\dd r}  {\mathbf \Omega}(r)
 =  \frac{i}{r}\,\mathbf{G}(r){\mathbf \Omega}(r), \quad r\ge r_1;
 \quad 
{\mathbf \Omega}(r_1) 
=
\begin{pmatrix}
 {\mathbf I} 
\\
- i {\mathbf Z}\o(r_1)
\end{pmatrix}. 
 \eeq
 Although ${\mathbf \Omega}(r) $ does not describe the complete wave field it is sufficient to determine the impedance, since
 \beq{296}
{\mathbf \Omega}(r) =  \begin{pmatrix}
{\mathbf \Omega}_1 
\\
{\mathbf \Omega}_2  
\end{pmatrix}
= {\mathbf M}(r,r_1)
\begin{pmatrix}
 {\mathbf I} 
\\
- i {\mathbf Z}\o(r_1)
\end{pmatrix}
\quad 
\Rightarrow
\quad 
{\mathbf Z}\o(r) = i {\mathbf \Omega}_2(r){\mathbf \Omega}_1^{-1}(r) ,
 \eeq
  for $r> r_1$.
The value of $r$ at which one switches from the differential Riccati equation  to the matricant based solution is a free parameter, and arbitrary as long as it is lies below the first singularity in the impedance.  This can be estimated from the separable solutions in \S\ref{sec7} as $ r_* \big(\omega^2 s_{max}^2-k_z^2\big)^{1/2} \sim  1$ where $s_{max}$ is the largest plane wave slowness at $r=0$.

\subsection{Discussion}\label{sec4.4}

We have described  two principal ways for finding the solid-cylinder impedance $\mathbf{Z}\o\left( r\right)$.  The Frobenius series method is  
summarized in  Eqs. \eqref{901} and \eqref{092}.  Taken together these equations provide a basis for calculating the solid-cylinder impedance for $n>0$ and arbitrary $r$ via  the Frobenius series solution.
The Riccati  equation method 
determines  $\mathbf{Z}\o\left( r\right) $  for arbitrary $n$ by integrating the differential Riccati equation \eqref{228}  subject to an initial condition defined by the central impedance $\mathbf{Z}\o_0$.  The Riccati approach  is strictly  
  valid  only  for   $r$  less than  the first singularity of the solid-cylinder impedance. 

The initial value ${\mathbf Z}_0$ can be evaluated from  \eqref{092}$_2$ or by other methods discussed in \S\ref{sec5}.  For $n=0$  the form of ${\mathbf Z}_0^{(0)}$ is explicit (eq. \eqref{-47} below) and $\mathbf{Z}^{\left(
0\right) }\left( r\right) $ may be determined by, for instance, integration of the Riccati equation discussed in \S\ref{4.3}.   
   The physical solution to the   initial value Riccati equation can be continued through and beyond the first   and  subsequent singularities  by using the matricant solution to generate  ${\mathbf Z}\o(r)$ as a conditional impedance.   Strictly speaking the practical value of the Riccati method is confined to the neighborhood of $r=0$.  The differential Riccati equation provides a regularization of the  system of equations \eqref{130} which are singular at $r=0$.  Once this singularity has been taken care of, there is no need to use the Riccati equation, particularly since the Riccati equation has its own singularities - in fact an infinite number of them.   {Note that satisfaction of the algebraic Riccati equation \eqref{25} is essential to ensure   regularization of the initial value problem \eqref{228} at $r=0$.   The  differential Riccati equation cannot generally   recover the 
   central impedance ${\mathbf Z}_0$ by `backward' integration to $r=0$ from some initial $r_0>0$,
    because the system possesses the same  ill-posed property observed with respect to eq.\ \eqref{814}, in this case associated with the fact that the nonphysical central impedance 
    $\mathbf{Z}_{\mathrm{np}}\left( 0\right) \ (=i\mathbf{L}_{2}%
\mathbf{A}_{2}^{-1}$ for $n>1)$ also solves \eqref{25}.  
   }

Both the Frobenius and Riccati methods generate an Hermitian  
solid cylinder impedance. 
Hermiticity of ${\mathbf Z}\o (r)$  is a consequence of the fact that it is built from   
 the triplet of  physical modes  which  produce zero normal fluxes both of their own and due to 
their cross-coupling.  Note that the nonphysical impedance $%
\mathbf{Z}_{\mathrm{np}}\left( r\right) =i\widehat{\mathbf{V}}_{2}\widehat{%
\mathbf{U}}_{2}^{-1}$ is Hermitian as well, which is similar to the case of a half-space; however, the physical and 
nonphysical impedances of a cylinder are generally no longer negative 
transpose of each other as they are for a half-space.     For $n>1$, the two impedances are related 
by
\beq{0344}
\mathbf{Z} \left( r\right)- \mathbf{Z}_{\mathrm{np}}\left( r\right) =i\big( \widehat{\mathbf{U}}_{1}\widehat{\mathbf{U}}_{2}^+\big)^{-1} , 
\nonumber
\eeq
with normalized $\widehat{\mathbf{U}}_{1}$, $\widehat{\mathbf{U}}_{2}$ on the right hand side, as follows from \eqref{885}$_2$ and \eqref{887}$_2$.  
The Hermitian nature of  ${\mathbf Z}\o(r)$, $r>0$ can also be viewed as a consequence of the fact that it solves the Riccati equation \eqref{228} with an Hermitian initial value, ${\mathbf Z}\o_0$. 
It is also noteworthy that neither the definition (\ref{901}) of $%
\mathbf{Z}\left( r\right) $ nor its Hermitian property    requires
 any specific normalization of the eigenvectors $\pmb{\gamma }_{\alpha }$ of 
$\mathbf{g}_{0}\left( 0\right) $ once they have been ordered into physical and nonphysical
triplets.

While the solid-cylinder impedance is  quite distinct in nature, it is 
in a  certain sense,  a conditional 
`one-point' impedance, for it depends on the initial condition at $r=0$. However, there is another, more essential, aspect, 
which actually sets $\mathbf{Z}( r)$ apart from the two-point impedance $\mathbf{Z}\left( r,r_0\right) $ and the 
general conditional impedance ${\mathbf z}\o (r)$.  
It is that $\mathbf{Z}\o\left( r,r_0\right) $ and ${\mathbf z}\o (r)$  involve all 6 linear independent partial solutions, whereas ${\mathbf Z}\o (r)$ involves only half of them and discards the
other half on the basis of certain partitioning at $r=0$ (physical/nonphysical). As a result, the Hermiticity of $\mathbf{Z}\left( r,r_0\right) $
  and ${\mathbf z}\o (r)$  and the Hermiticity
of ${\mathbf Z}\o (r)$ have different origins.    Hermiticity of both $\mathbf{Z}\left( r,r_0\right) $ and ${\mathbf z}\o (r)$  (the latter subject to Hermiticity of the initial condition) follows from $\div {\mathbf P}=0$    while 
Hermiticity of   $\mathbf{Z}\left( r \right)$ is a consequence of $P_r = 0$. 


\section{Properties of the central impedance matrix $\mathbf{Z}\o_0$ }\label{sec5}

 The central-impedance matrix depends only on the elastic moduli (and $n$) and is  
 simpler than the solid-cylinder impedance, its continuation away from $r=0$. At the same time, 
the value of the central-impedance is required {\it a priori} in order to calculate $\mathbf{Z}(r)$ using the Riccati equation \eqref{228}.    
In this section we describe some properties of $\mathbf{Z}\o_0$, develop procedures for its determination, and consider its behavior for large $n$.

\subsection{Integral formula for $ {\mathbf Z}\o_0$ with  $n>1$}\label{5.1}  

\subsubsection{The Lothe-Barnett integral method using the matrix sign function}

The surface impedance matrix $\mathbf{Z}(v)$, $v$ =velocity, plays a central
part in the theory of surface waves in an elastic homogeneous half-space. It
was first identified in that context by \citet{Ingebrigtsen69} and
 subsequently developed as a  crucial ingredient for proving the uniqueness and
the existence conditions for surface waves \cite{Lothe76,Barnett85}, see also \cite{Chadwick77,Ting96}.  The
central impedance matrix $\mathbf{Z}_{0}$ of a cylinder has a close
relationship to the static ($v=0$) surface impedance matrix. The similarity
allows us to use some of the considerable array of results for the latter.
Here we draw directly on the integral formalism for the surface impedance of
a half-space first outlined by \citet{Barnett73,Barnett74} and then presented in full by \citet{Lothe76} 
and \citet{Chadwick77}. In this subsection we show how this formalism
can be modified to describe the cylinder central impedance $ {\mathbf Z}\o_0$ for 
 $n>1$, and we discuss the exceptional cases  $n=0,1.$

Assume $n>1,$ so that the eigenspectrum of $\mathbf{g}_{0} ( 0 )$ lies on either side of the imaginary axis in accordance with  eq. \eqref{883}.  
The restriction to $n>1$ will be clarified below. 
The  matrix sign function  is then uniquely defined  
\beq{273}
\sign \mathbf{g}_{0} ( 0 ) = \mathbf{g}_{0} ( 0 ) \, \big( \mathbf{g}_{0}^2 ( 0 ) \big)^{-1/2} ,  
\eeq
where the principle branch of the square root function with branch cut on the negative real axis is understood;  $z = (z^2)^{1/2}\, \sign z$ with  $ \sign z =+1 (-1)$ if $\Re z >0 (<0)$.
As a result the sign matrix satisfies 
\begin{equation}
\big( \sign  \mathbf{g}_{0} ( 0 )\big) \pmb{\gamma }_{\alpha }=\pm \pmb{\gamma }_{\alpha }\ \mathrm{for}\ \Re %
\lambda _{\alpha }\gtrless 0.   \label{7.15}
\end{equation}%
 The  matrix sign function  was first introduced by Roberts \cite{Roberts:1980:LMR} 
 as a means of solving algebraic Riccati equations, and has  become a standard matrix function, see \cite{Kenney:1995:MSF,Higham:2008:FM} 
 for   reviews; the simple relation \eqref{273} was first noted in  \cite{Higham:1994:MSD}. 
Using  the spectral decomposition defined by    
  the matrix of eigenvectors $ \mathbf{\Gamma }_{0}$ yields 
\bal{332}
  & \mathbf{g}_{0}( 0 )  = \mathbf{\Gamma }_{0} \mathbf{\Lambda }_0 \mathbf{\Gamma }_{0}^{-1}
   \quad \Rightarrow \quad
   \sign  \mathbf{g}_{0}( 0 )
   = \mathbf{\Gamma }_{0}  \big(\sign \mathbf{\Lambda }_0\big) \mathbf{\Gamma }_{0}^{-1}
  \quad \text{with}
  \nonumber \\
    & \mathbf{\Lambda }_0 = 
   \begin{pmatrix}
\mathrm{diag} \big(\lambda _{\alpha }\big)   & {\mathbf{0}} \\ & \\
{\mathbf{0}} & \mathrm{diag} \big(-\lambda _{\alpha }^{\ast} \big)
\end{pmatrix} 
\quad \Rightarrow \quad
\sign \mathbf{\Lambda }_0 
= 
   \begin{pmatrix}
\mathbf{I}   & {\mathbf{0}} \\ & \\
{\mathbf{0}} & - \mathbf{I} 
\end{pmatrix} .
\end{align}
The explicit structure of the sign matrix follows from  the  normalization condition \eqref{884} and the submatrices defined in \eqref{-332}$_2$, 
\bal{7.21}
  \sign  \mathbf{g}_{0}  (0)
&=   \mathbf{\Gamma }_{0}  \big(\sign \mathbf{\Lambda }_0\big) \mathbf{T} \mathbf{\Gamma }_{0}^+\mathbf{T} 
= 
\begin{pmatrix}
\mathbf{S} & i\mathbf{H} \\ 
i\mathbf{B} & -\mathbf{S}^{+}%
\end{pmatrix} \quad
\text{with}
\nonumber \\
\mathbf{S}=2\mathbf{A}_{1}\mathbf{L}_{2}^{+}-\mathbf{I}
&= \mathbf{I} - 2\mathbf{A}_{2}\mathbf{L}_{1}^{+},
\quad
\mathbf{H}=-2i\mathbf{A}_{1}\mathbf{A}_{2}^{+} = \mathbf{H}^+,
\quad \mathbf{B}=-2i\mathbf{L}_{1}%
\mathbf{L}_{2}^{+}= \mathbf{B}^+.  
\end{align}%
Additional relations are obtained from the involutory property of the sign matrix function, 
\begin{equation}
\big(  \sign  \mathbf{g}_{0}  (0)\big) ^{2}=\mathbf{%
I}_{6}
\quad \Rightarrow \quad
\mathbf{S}^{2}-\mathbf{HB=I,\ SH}=\left( \mathbf{SH}\right) ^{+},\ 
\mathbf{BS}=\left( \mathbf{BS}\right) ^{+}.  \label{7.16}
\end{equation}%

The connection with  Barnett and Lothe's theory is established  via the integral expression for the matrix sign function \cite{Roberts:1980:LMR,Higham:2008:FM}
\beq{33.21}
  \sign  \mathbf{g}_{0} (0) = 
  \frac2{\pi}\mathbf{g}_{0}(0) \int_0^\infty \dd t\, \big( t^2 \mathbf{I}+ \mathbf{g}_{0}^2(0)
  \big)^{-1}. 
\eeq
A simple change of integration variable and separation into partial fractions yields  $\sign  \mathbf{g}_{0}(0)$
as an  averaged matrix,  
\beq{-55-}
 \sign  \mathbf{g}_{0} (0)=\frac{1}{%
\pi }\int\limits_{0}^{\pi }\dd 
\phi \, \mathbf{g}_{0}^{ ( \phi  ) }
\equiv \big\langle  \mathbf{g}_{0}^{ ( \phi  ) } \big\rangle ,
\eeq
where 
\beq{-4235}
\mathbf{g}_{0}^{ ( \phi  ) }
=\big( \cos \phi\, \mathbf{I}-i\sin \phi\, \mathbf{g}_{0}(0)
\big) ^{-1}\big( \cos \phi \, \mathbf{g}_{0}(0)-i\sin \phi \, \mathbf{I}\big).
\eeq
The latter can be simplified by noting   
\beq{-66-}
  \mathbf{g}_{0} (0)=\big( \mathbf{A}_0 -  \mathbf{B}_0\big)^{-1}
 \widehat{\mathbf{J}}
 \big( \mathbf{A}_0 +  \mathbf{B}_0\big),
 \eeq
 with 
 \beq{-54-}
 \widehat{\mathbf{J}} = i 
 \begin{pmatrix}
 \mathbf{0}  & \mathbf{I} \\ 
- \mathbf{I} & \mathbf{0}%
\end{pmatrix} ,
\quad
\mathbf{A}_0 -  \mathbf{B}_0 = 
 \begin{pmatrix}
-\widehat{\mathbf{Q}}  & \mathbf{0} \\ 
-i\widetilde{\mathbf{R}}^+  & \mathbf{I}%
\end{pmatrix},
\quad
\mathbf{A}_0 +  \mathbf{B}_0 = 
\begin{pmatrix}
-\widetilde{\mathbf{T}}  & \mathbf{0} \\ 
-i\widetilde{\mathbf{R}}  & \mathbf{I}%
\end{pmatrix}.
 \eeq
Therefore,  
 \beq{-67-}
 \mathbf{g}_{0}^{ ( \phi  ) }
=  
 \big( \mathbf{A}_0 - e^{i2\phi \widehat{\mathbf{J}}} \mathbf{B}_0\big)^{-1}
 \widehat{\mathbf{J}}
 \big( \mathbf{A}_0 + e^{i2\phi \widehat{\mathbf{J}}} \mathbf{B}_0\big), 
 \eeq
 and defining, by analogy with \eqref{-54-},
\beq{033-}
\mathbf{A}_0 - e^{i2\phi \widehat{\mathbf{J}}} \mathbf{B}_0 
=  \begin{pmatrix}
-\widehat{\mathbf{Q}}_\phi  & \mathbf{0} \\ 
-i\widetilde{\mathbf{R}}^+_\phi  & \mathbf{I}%
\end{pmatrix}
,
\quad
 \mathbf{A}_0 + e^{i2\phi \widehat{\mathbf{J}}} \mathbf{B}_0 
 =
\begin{pmatrix}
-\widetilde{\mathbf{T}}_\phi  & \mathbf{0} \\ 
-i\widetilde{\mathbf{R}}_\phi  & \mathbf{I}%
\end{pmatrix}
,
 \eeq
 then the matrix $\mathbf{g}_{0}^{( \phi ) }$ can be expressed in exactly the  same structural form as 
 $\mathbf{g}_{0}(0)$ in terms of $3\times 3$ matrices, as
 \beq{7.12}
\mathbf{g}_{0}^{( \phi ) }= 
\begin{pmatrix} 
-\widehat{\mathbf{Q}}_{\phi }^{-1}\widetilde{\mathbf{R}}_{\phi } & -i%
\widehat{\mathbf{Q}}_{\phi }^{-1}\  \\ 
i\big( \widetilde{\mathbf{T}}_{\phi }-\widetilde{\mathbf{R}}_{\phi }^{+}%
\widehat{\mathbf{Q}}_{\phi }^{-1}\widetilde{\mathbf{R}}_{\phi }\big) & 
\widetilde{\mathbf{R}}_{\phi }^{+}\widehat{\mathbf{Q}}_{\phi }^{-1}%
\end{pmatrix} ,  
\eeq
where the   $\pi -$periodic  submatrices are
 \bal{7.121}
\widehat{\mathbf{Q}}_{\phi }=\widehat{\mathbf{Q}}_{\phi }^{+}
&=\cos^2 \phi \, 
\widehat{\mathbf{Q}}+\sin ^{2}\phi \,\widetilde{\mathbf{T}}+i\sin \phi \cos
\phi \big( \widetilde{\mathbf{R}}-\widetilde{\mathbf{R}}^{+}\big) =%
\widetilde{\mathbf{T}}_{\phi +\pi /2}, \nonumber 
\\ 
\widetilde{\mathbf{T}}_{\phi }=\widetilde{\mathbf{T}}_{\phi }^{+}
&=\cos^{2}\phi\, \widetilde{\mathbf{T}}+\sin ^{2}\phi \, \widehat{\mathbf{Q}}-i\sin
\phi \cos \phi \big( \widetilde{\mathbf{R}}-\widetilde{\mathbf{R}}%
^{+}\big) =\widehat{\mathbf{Q}}_{\phi +\pi /2}, 
\\ 
\widetilde{\mathbf{R}}_{\phi }&=\cos ^{2}\phi \,\widetilde{\mathbf{R}}+\sin^{2}\phi \,\widetilde{\mathbf{R}}^{+}+i\sin \phi \cos \phi \big( \widehat{%
\mathbf{Q}}-\widetilde{\mathbf{T}}\big) =\widetilde{\mathbf{R}}_{\phi +\pi
/2}^{+}. \nonumber 
\end{align}%
The submatrices defined in eq. \eqref{7.21} therefore have  alternative integral expressions, from \eqref{-55-} 
and \eqref{7.121},  
\bal{7.13}
\mathbf{S=}-\big\langle\widehat{\mathbf{Q}}_{\phi
}^{-1}\widetilde{\mathbf{R}}_{\phi }\big\rangle ,
\ \mathbf{H=}-\big\langle\widehat{\mathbf{Q}}_{\phi }^{-1}\big\rangle \ 
,
\ \mathbf{B}=\big\langle 
\widetilde{\mathbf{T}}_{\phi }-\widetilde{\mathbf{R}}_{\phi }^{+}\widehat{%
\mathbf{Q}}_{\phi }^{-1}\widetilde{\mathbf{R}}_{\phi }  \big\rangle
.
\nonumber 
\end{align}%

Crucially, $\widehat{\mathbf{Q}}_{\phi } $ is positive definite  for $n>1$.   In order to see this, first note the obvious positive definiteness  $\widehat{\mathbf{Q}}_{\phi }>0$ if  $\sin \phi =0$.  For $\sin \phi \ne 0$  we have 
\beq{77}
\widehat{\mathbf{Q}}_\phi = 
-\sin^{2}\phi \mathbf{\Lambda } \big( -i\cot \phi \big)
\quad \text{where}\quad
{\mathbf \Lambda} (\lambda ) \equiv
\lambda^2 {\mathbf{\widehat{Q}}}_0 + \lambda (  \widetilde{\mathbf R}_0 - \widetilde{\mathbf R}_0^+)
- \widetilde{\mathbf T}_0, 
\eeq
which is positive definite by virtue of the fact that 
$\det \mathbf{\Lambda }\left( \lambda \right) =0$ does not admit pure
imaginary roots for $\lambda $ \citep[\S 3.2.1]{Shuvalov03a}.
By the above arguments, the matrices $\mathbf{H}$ and $\mathbf{B}$ are
negative and positive definite, respectively.  Consequently, $\mathbf{H}$
and $\mathbf{B}$ are invertible, and so  the identity  
\eqref{7.16}$_2$ implies that  the matrices $\mathbf{I-S}^{2}
$ and hence $\mathbf{I-S}^{+2}$ are also  invertible. 
Note that the  positive definiteness of  $\mathbf{B}$ confirms that $\det \mathbf{B}%
\neq 0$\ for $n>1$, which is when there is no stress-free modes.  
The cases $n=0$, $1$ are discussed in \S \ref{sec5111}.

\subsubsection{The impedance matrices $\mathbf{Z}_{0}$ and $\mathbf{Z}_{0\mathrm{np}}$} 

We are now in a position to express the impedance matrices in terms of the integrals. 
As before, we set  $\alpha =1,2,3$ and $\alpha =4,5,6$ for the
physical and nonphysical triplets, respectively.  Inserting 
$\mathbf{L}_{1}=-i\mathbf{Z}_{0}\mathbf{A}_{1},$ $\mathbf{L}_{2}=-i\mathbf{Z}_{0\mathrm{np}}%
\mathbf{A}_{2}$ in 
\eqref{-332}$_2$ and using the same argument as in \cite{Lothe76} to maintain that $\det 
\mathbf{A}_{1,2}\neq 0$ implies 
\beq{4-4}
 \mathbf{\Gamma }_{0}
 = \begin{pmatrix}
\mathbf{I}& \mathbf{I}
\\
-i \mathbf{Z}_{0} & -i\mathbf{Z}_{0\mathrm{np}}
\end{pmatrix}
\begin{pmatrix}
\mathbf{A}_1 & \mathbf{0}
\\
 \mathbf{0} & \mathbf{A}_2
\end{pmatrix}, 
 \eeq
which, together with  eqs. \eqref{332} \eqref{7.21}, yields the matrix identity 
\beq{-445}
\begin{pmatrix}
\mathbf{S} & i\mathbf{H} \\ 
i\mathbf{B} & -\mathbf{S}^{+}%
\end{pmatrix}  \begin{pmatrix}
\mathbf{I}& \mathbf{I}
\\
-i \mathbf{Z}_{0} & -i\mathbf{Z}_{0\mathrm{np}}
\end{pmatrix} = 
 \begin{pmatrix}
\mathbf{I}& -\mathbf{I}
\\
-i \mathbf{Z}_{0} & i\mathbf{Z}_{0\mathrm{np}}
\end{pmatrix} .
\eeq
The first line yields explicit expressions for the impedances $\mathbf{Z}_{0}$ and $%
\mathbf{Z}_{0\mathrm{np}}$, 
\begin{equation}
\mathbf{\mathbf{Z}}_{0}=\mathbf{H}^{-1}-\mathbf{H}^{-1}\mathbf{S,\qquad 
\mathbf{Z}%
}_{0\mathrm{np}}=-\mathbf{H}^{-1}-\mathbf{H}^{-1}\mathbf{S},  \label{7.22}
\end{equation}%
and the second line gives the equivalent expressions 
\begin{equation}
\mathbf{\mathbf{\mathbf{Z}}}_{0}=\mathbf{-\left( \mathbf{I+S}^{+}\right) }%
^{-1}\mathbf{\mathbf{B=-B}\left( \mathbf{I+S}\right) }^{-1}\mathbf{,\ 
\mathbf{Z}}_{0\mathrm{np}}=\left( \mathbf{I-S}^{+}\right) ^{-1}\mathbf{B=B}%
\left( \mathbf{I-S}\right) ^{-1}\mathbf{.}  \label{7.221}
\end{equation}%
Hence,  $\mathbf{Z}_{0}$ and $\mathbf{Z}_{0\mathrm{%
np}}$ are Hermitian by virtue of  (\ref{7.22}) and (\ref{7.16}),
\begin{equation}
\mathbf{\mathbf{Z}}_{0}=\mathbf{\mathbf{Z}}_{0 }^{+},\qquad
\mathbf{Z}_{0\mathrm{np}}=\mathbf{Z}_{0\mathrm{np}}^{+}.  \label{7.23}
\end{equation}%
Using  (\ref{7.22}) and (\ref{7.21}) implies 
\begin{equation}
\mathbf{\mathbf{Z}}_{0}-\mathbf{Z}_{0\mathrm{np}}=2\mathbf{H}^{-1}=i\left( 
\mathbf{A}_{1}\mathbf{A}_{2}^{+}\right) ^{-1}\left( =-i\left( \mathbf{A}_{2}%
\mathbf{A}_{1}^{+}\right) ^{-1}\right)   \label{7.24}
\end{equation}%
in agreement with the formula in \S \ref{sec4.4}.

The representation (\ref{7.22}) of the physical and non-physical central
impedances is similar to that of, respectively, the non-physical and physical
half-space impedance $\mathbf{\mathbf{Z}}\left( v\right) $ in \cite{Lothe76}; however, the 
two addends in the right members of (\ref{7.22}) are generally not the real and imaginary
parts of the impedance as  is the case in the  expressions of \cite{Lothe76}. Consequently,  the non-physical central impedance is not minus transpose of the
physical one, unlike  the half-space impedance where $\mathbf{Z}\left( v\right) =-\mathbf{Z}_{\mathrm{np}}^{\mathrm{T}}\left( v\right) $ \cite{Lothe76}.
Another noteworthy difference is that the expressions similar to (\ref{7.221}%
) are not unreservedly valid for the dynamical $\mathbf{Z}\left( v\right) $
because the analogue of $\mathbf{B}$ has zero determinant at the Rayleigh
speed.

The above results can in the main be obtained by following the alternative
method of deriving the Lothe-Barnett integral formalism which was proposed
by Mielke and Fu in \cite{Mielke04}. 

\subsection{Remarks on $n=0$, $1$} \label{sec5111}
Let us now discuss the implication of the special cases $n=0$, $1$ in the
present context. Occurrence of a non-semisimple 
$\mathbf{g}_{0}\left(0\right) $ due to a pair of degenerate eigenvalues 
$\lambda ^{\left(0,1\right) }=0$, which are split between physical and nonphysical triplets,
makes the cases $n=0$, $1$ tantamount to the limiting state $v=\widehat{v}$ of
the elastodynamic problem  for a half-space \cite{Lothe76,Barnett85,Chadwick77} (more specifically, to the
so-called exceptional limiting state in view of the zero-traction mode
corresponding to $\lambda ^{\left( 0,1\right) }=0$). The Lothe-Barnett  
integral formalism on the whole is well-defined at $v<\widehat{v}.$ Any 
difficulties occurring at $v=\widehat{v}$ are due to the
non-integrable divergence acquired at $v=\widehat{v}$ by the angularly
varying Stroh matrix $\mathbf{N}\left( \widehat{v},\phi \right) ,$ which is
a counterpart of $\mathbf{g}_{0}^{\left( \phi \right) }$.  
A similar exception  arises with  $\mathbf{g}_{0}^{\left( \phi \right) }$ for $n=0$, $1$ due to  $\widehat{ \mathbf{Q}}_{\phi }^{-1}.$  The   argument underlying positive
definiteness of $\widehat{\mathbf{Q}}_{\phi }$ for $n>1$  no longer
applies for the cases $n=0$, $1$ which admit rigid-body motion.  This can be ascribed to the fact that  $\det \mathbf{\pmb \kappa }^{\left( 0,1\right) }=in\left(
n^{2}-1\right) =0$ 
so that $\det 
\mathbf{\Lambda }\left( \lambda \right) =0$ 
has the root $\lambda ^{\left( 0,1\right) }=0$, where  
$\mathbf{\Lambda }\left( \lambda \right) $ is given
by \eqref{77}.   Thus $\widehat{\mathbf{Q}}%
_{\phi }$ for $n=0$, $1$ is positive semi-definite, with $\det \widehat{\mathbf{%
Q}}_{\phi }=0$ at $\cos \phi =0.$  That is why the cylinder's version of the
integral formalism cannot generally be extended to the cases $n=0$, $1$. The
exception when this is yet possible is the case $n=0$ for a cylindrically
monoclinic material with the symmetry plane orthogonal to the $z$-axis. 
This case simplifies due to the simultaneous occurrence of 
 $\mathbf{e}_{z}$ as the null vector of $\pmb{\kappa 
}^{\left( 0\right) }$ and of the uncoupling of the $zz$-components. Hence
the upper 2$\times $2 blocks of the integral-formalism relations remain
valid. Such a state of affairs also has a direct analogy with the theory of
surface impedance in half-space, namely, with the case of a symmetrical
sagittal plane, which is when the in-plane modes are unaffected by the
limiting state $\widehat{v}_{SH}$ of the uncoupled shear-horizontal mode 
\cite{Chadwick90,Barnett92}. 
 A careful remark is in order
regarding Eq. (\ref{7.221}). For $n=0,$ the rigid-body displacement
corresponding to $\lambda ^{\left( 0\right) }=1$ and parallel to $\mathbf{e}%
_{\theta }$ is the null vector of $\mathbf{B}$ and the eigenvector of $%
\mathbf{S}$ with the eigenvalue $\left\langle \lambda ^{\left( \phi \right)
}\right\rangle =1.$ Hence the upper 2$\times $2 blocks of $\mathbf{B}$ and $%
\mathbf{I-S,\ I-S}^{+}$ are singular, whereas those of $\mathbf{I+S,\ I+S}^{+}
$ are not. Thus (\ref{7.221}) is not valid for $\mathbf{Z}_{0\mathrm{np}}$
even for the monoclinic $n=0$ case.

Finally, it needs to be added that  analysis  of the
half-space integral formalism as $v\rightarrow \widehat{v}$ \cite{Lothe76,Chadwick77,Ting96,Fu02} 
shows that although the formalism diverges on the whole in this limit, the
integral expressions for the surface impedance $\mathbf{Z}\left( v\right) $
remain well-defined at $v=\widehat{v}.$ This asymptotic property of $%
\mathbf{Z}\left( v\right) $ is not however directly relevant to the central
impedance $\mathbf{Z}_{0}$ of a cylinder, since the diverging cases $n=0$, $1$
cannot be approached 'continuously in $n$'. A more appropriate treatment is
either asymptotic analysis of $\mathbf{Z}\left( r\right) \rightarrow \mathbf{%
Z}_{0}$ as $r\rightarrow 0$ or else other, explicit, methods of deriving $%
\mathbf{Z}_{0}$ for $n=0$, $1$, see \S \ref{5.3}.

\subsection{Definiteness of $\mathbf{Z}_{0}$ for 
$n>1$ and  semi-definiteness for $n=0,1$}\label{sec5.25}
It has been noted above that the structure of the physical and non-physical
central impedances (\ref{7.22}) resembles that of, respectively, the
non-physical and physical surface impedances \cite{Lothe76} for a half-space. This
suggests the inverse correspondence of their sign properties. We will
outline a formal proof. Similarly to the Lothe-Barnett theory for a surface
impedance, the insight does not follow from the integral formalism but
relies instead on  static energy considerations.

As before, assume first that $n>1$. The central impedance is related to $%
\mathbf{g}_{0}\left( 0\right) $ independent of $\omega $ and $k_{z},$
therefore we can invoke the 2D static solution 
$\pmb{\mathcal{N}}\left(r\right) =\mathbf{\Gamma }\, \mathrm{diag}\left( r^{\lambda _{\alpha }}\right) $ ($\omega =0,~k_{z}=0$ is tacit below). The time averaged energy associated with
these solutions  is 
\citep[Eq.\ (21)]{Shuvalov03a}
\begin{equation}
W=-\frac{i}{8r}\frac{\mathrm{d}}{\mathrm{d}r}\left( \mathbf{U}^{+ }%
\mathbf{V} -\mathbf{V}^{+ }\mathbf{U}\right) .
\label{10.18}
\end{equation}%
Inserting the physical solutions with  eigenvalues $\Re \lambda _{\alpha }>0$ 
and eigenvectors 
$\pmb{\gamma}_{0\alpha }=\left( \mathbf{a}_{\alpha },\mathbf{l}_{\alpha }\right) ^{%
\mathrm{T}}$ ($\alpha =1,2,3$) of $\mathbf{g}_{0}\left( 0\right) $ from 
 \S \ref{sec4.1.1}
and using  the central impedance $\mathbf{Z}_{0}=\mathbf{Z}%
_{0}^{+}$ leads to%
\begin{equation}
\int_{r_{1}}^{r_{2}}Wr
\mathrm{d}r =- \frac{1}{4}\sum\limits_{\alpha=1}^3 \left( r_{2}^{2\Re \lambda _{\alpha }}-r_{1}^{2\Re \lambda _{\alpha }}\right) \mathbf{a}%
_{\alpha }^{\ast }\mathbf{Z}_{0}\mathbf{a}_{\alpha }>0,\ \forall \, r_2 >r_1.  \label{10.20}
\end{equation}%
Hence $\mathbf{Z}_{0}$ for $n>1$ is negative definite. The same
consideration using the non-physical solutions with $\Re \lambda
_{\alpha }<0,$ $\alpha =4,5,6,$ implies that $\mathbf{Z}_{0\mathrm{np}}$ for 
$n>1$ is positive definite. As expected, this is opposite to the sign
properties of the physical and non-physical surface impedances $\mathbf{Z}\left(
v\right)$, $\mathbf{Z}_{\mathrm{np}}\left( v\right) $ for the static limit $v=0$.

In the case $n=1,$ the above proof applies unchanged for the physical $%
\mathbf{Z}_{0}^{\left( 1\right) }$ except that it is negative semi-definite
due to the presence of the rigid-body motion mode. In the case $n=0,$ the
same conclusion of negative semi-definite $\mathbf{Z}_{0}^{\left( 0\right) }$
follows from an explicit calculation of $\mathbf{Z}_{0}^{\left( 0\right)
}=\lim_{r\rightarrow 0}\mathbf{Z}^{\left( 0\right) }\left( r\right) $
presented in \S \ref{5.3.2}.
It is noted that the rigid-body displacements causing 
\beq{111} \det \mathbf{Z}_{0}=0 \quad \text{for }n=0,1,
\eeq
are related to the existence of low frequency (long wavelength) guided waves
in rods: longitudinal, torsional ($n=0$) and flexural ($n=1$), see \S\ref{4.1.2}
and \cite{Shuvalov03a}.

\subsection{Explicit expressions for the central impedance matrix}\label{5.3}

 In this subsection we develop other procedures for determining ${\mathbf Z}_0$, including for the special  cases $n=0$ and $n=1$. 

\subsubsection{$\mathbf{Z}\o_0$ for $n>0$}\label{5.3.1}

The central impedance $\mathbf{Z}_{0}$ for $n>0$ is defined by any  of the
optional relations (including \eqref{092}$_2$) that may be written similarly to
\eqref{901} as%
\beq{2222}
\mathbf{l}_{\alpha }=-i\mathbf{Z}_{0}\mathbf{a}_{\alpha }\ \ \ \left( \alpha
=1,2,3\right) 
\quad \Leftrightarrow \quad \mathbf{l}=-i\mathbf{Z}_{0}\mathbf{a}%
\quad\Leftrightarrow \quad\mathbf{Z}_{0}=i\mathbf{L}_{1}\mathbf{A}_{1}^{-1},
\eeq
where $\pmb{\gamma }=\left( \mathbf{a},\mathbf{l}\right) ^{\mathrm{T}}$
is an arbitrary superposition of the physical eigenvectors $\pmb{\gamma 
}_{\alpha }=\left( \mathbf{a}_{\alpha },\mathbf{l}_{\alpha }\right) ^{%
\mathrm{T}}$ of $\mathbf{g}_{0}\left( 0\right) $ with $\alpha =1,2,3$. The
matrices $\mathbf{A}_{1}=\left\Vert \mathbf{a}_{1},\mathbf{a}_{2},\mathbf{a}%
_{3}\right\Vert $ and $\mathbf{L}_{1}=\left\Vert \mathbf{l}_{1},\mathbf{l}%
_{2},\mathbf{l}_{3}\right\Vert $ may be related to one another using identities such as $(24) $ and $(26)$  of
\cite{Shuvalov03a},
\beq{73} 
{\mathbf l}_\alpha = i ( \lambda_\alpha {\mathbf{\widehat{Q}}}_0 + \widetilde{\mathbf R}_0 ){\mathbf a}_\alpha 
\quad (\alpha=1,2,3), 
\nonumber
\eeq
implying  
 \beq{731}
{\mathbf L}_1 = i ( {\mathbf{\widehat{Q}}}_0 {\mathbf A}_1{\pmb \lambda}+ \widetilde{\mathbf R}_0 {\mathbf A}_1 )  ,
\quad \text{ where}\quad
{\pmb \lambda} =\diag (\lambda_1, \lambda_2, \lambda_3).
\eeq
The eigenvectors ${\mathbf a}_\alpha $ $(\alpha =1,2,3)$    are null vectors of
${\mathbf \Lambda} (\lambda_\alpha)$, see. eq. \eqref{77},  
and consequently,
\beq{78}
  {\mathbf{\widehat{Q}}}_0 {\mathbf A}_1{\pmb \lambda}^2 +   (   \widetilde{\mathbf R}_0 - \widetilde{\mathbf R}_0^+ ){\mathbf A}_1{\pmb \lambda}
- \widetilde{\mathbf T}_0 {\mathbf A}_1={\mathbf{0}} .
\eeq
Equations \eqref{731} and \eqref{78}   provide a pair of  expressions for the central impedance, each in terms of the displacement eigenvector matrix only,
\beq{74}
{\mathbf Z}\o_0 = - \widetilde{\mathbf R}_0 - {\mathbf{\widehat{Q}}}_0 {\mathbf A}_1{\pmb \lambda} {\mathbf A}_1^{-1}
 =   - \widetilde{\mathbf R}_0^+  - \widetilde{\mathbf T}_0  {\mathbf A}_1{\pmb \lambda}^{-1} {\mathbf A}_1^{-1}.
\eeq
The freedom afforded by these simultaneous identities  will prove to be  useful when  material symmetry reduces the matrix size to $2\times 2$, see \S\ref{7.1.1}.

\subsubsection{The central impedance for $n=0$}\label{5.3.2} 

The  algebraic Riccati equation \eqref{25}  for $n=0$ leads to  a constructive solution for ${\mathbf Z}^{(0)}$, which must satisfy
\beq{-45}
  \bigg\{ {\mathbf Z}^{(0)}_0 
+ \begin{pmatrix}
c_{12} & c_{26}  & c_{25} 
\\ 
-c_{16} & -c_{66}  & -c_{56}
\\ 
0 & 0 & 0
\end{pmatrix}
\bigg\} {\mathbf{\widehat Q}}_0^{-1} 
\bigg\{ {\mathbf Z}^{(0)}_0 
+ \begin{pmatrix}
c_{12} & -c_{16}  & 0
\\ 
c_{26} & -c_{66}  & 0
\\ 
c_{25} & -c_{56} & 0
\end{pmatrix}
\bigg\}  
+ \begin{pmatrix}
-c_{22} & c_{26}  & 0
\\ 
c_{26} & -c_{66}  & 0
\\ 
0 & 0 & 0
\end{pmatrix}
= {\mathbf{0}} .\nonumber
\eeq
Noting that 
\beq{-46}
 \begin{pmatrix}
0 & 0  & 0 
\\ 
c_{16} & c_{66}  & c_{56}
\\ 
0 & 0 & 0
\end{pmatrix}
 {\mathbf {\widehat Q}}_0^{-1} 
 = 
{\mathbf {\widehat Q}}_0^{-1} 
\begin{pmatrix}
0 & c_{16}  & 0
\\ 
0 & c_{66}  & 0
\\ 
0 & c_{56} & 0
\end{pmatrix}
 = 
\begin{pmatrix}
0 & 0 & 0
\\ 
0 & 1 & 0
\\ 
0 &0 & 0
\end{pmatrix}, \nonumber
\eeq
it is clear that the solution of the algebraic Riccati equation is of the form 
\beq{-47}
{\mathbf Z}^{(0)}_0  = 
\begin{pmatrix}
z^{(0)} & 0 & 0
\\ 
0 & 0 & 0
\\ 
0 &0 & 0
\end{pmatrix},
\eeq
where the scalar $z^{(0)}$ satisfies a quadratic equation
\beq{-48}
\big( z^{(0)} {\mathbf e}_r + {\mathbf p}\big)^T
{\mathbf {\widehat Q}}_0^{-1}
\big( z^{(0)} {\mathbf e}_r + {\mathbf p}\big)
-c_{22} = 0,
\quad \text{with   } 
{\mathbf p}^T
= \big( c_{12}  ,\, c_{26} , \,   
c_{25}\big) . \nonumber
\eeq
The physical root   must have negative real part in order to be consistent with eq. \eqref{-3-} below, implying 
\beq{3-}
z^{(0)} =\frac1{\big(\widehat{Q}^{-1}_0\big)_{11} }\big[ -q_1 - \sqrt{ 
q_1^2 + \big( c_{22} - {\mathbf p}^T {\mathbf q}\big)\big(\widehat{Q}^{-1}_0\big)_{11} } \big],
\quad \text{where } 
{\mathbf q} =  {\mathbf {\widehat Q}}_0^{-1}  {\mathbf p}.
\eeq

An alternative method is 
to take the limit $r\rightarrow 0$ of the solution 
for $\pmb{\mathcal{N}}$ of \cite[Eq. A4]{Shuvalov03a}.   The result is again eq.\ \eqref{-47} 
where, by definition,  $z^{\left( 0\right) }$ is $(i^2 r)$ times the ratio of radial components of the traction and displacement of the eigenvector $\pmb{\gamma }^{\left(
0\right) }$ of $\mathbf{g}_{0} (0)$ corresponding to its
eigenvalue $\lambda =\kappa$.   These were   found by \citet{Ting96a}, from which $z^{\left( 0\right) }$ follows  as
\beq{-3-}
\ z^{\left( 0\right) }=-\, \frac{W+\sqrt{YQ}}{%
c_{55}c_{66}-c_{56}^{2}},
\eeq
where $Q = \det \mathbf{\widehat{Q}}_0$ and, using the notation of \cite{Ting96a},
\beq{008}
{W}=\det\begin{pmatrix} 
c_{12} & c_{26} & c_{25} \\ 
c_{16} & c_{66} & c_{56} \\ 
c_{15} & c_{56} & c_{55}%
\end{pmatrix} ,
\qquad
 {Y}=\det\begin{pmatrix} 
c_{22} & c_{26} & c_{25} \\ 
c_{26} & c_{66} & c_{56} \\ 
c_{25} & c_{56} & c_{55}%
\end{pmatrix}.\nonumber
\eeq
Equivalence of the expressions \eqref{3-} and \eqref{-3-} follow from identities such as $q_1=W/Q$ and $(
c_{22} - {\mathbf p}^T {\mathbf q} ) \big(\widehat{Q}^{-1}_0\big)_{11} 
=(QY - W^2)/Q^2$.   Note that $QY > W^2$ \cite[Eq.\ (B2)]{Ting96a}.

\subsubsection{ $ {\mathbf Z}\o_0$ for $n=1$ } \label{5.3.3}

The physical triplet of eigenvalues $\lambda _{\alpha }^{(
1) }$ and eigenvectors $\pmb{\gamma }_{\alpha }^{( 1)
}=\big( \mathbf{a}_{\alpha }^{\left( 1\right) },\mathbf{l}_{\alpha
}^{\left( 1\right) }\big) ^{\mathrm{T}}$ 
of $\mathbf{g}_{0} \left( 0\right) $ ($\alpha =1,2,3$) includes $\lambda ^{\left(
1\right) }=0$.  It corresponds to a rigid-body rotation
about the $z-$axis with displacement vector $\mathbf{a}^{\left( 1\right) }=\left( 1,i,0\right) ^{\mathrm{T%
}}$ and zero traction $\mathbf{l}^{\left( 1\right) }=\mathbf{0}$, 
see \cite[Eq. (52)]{Shuvalov03a}. Hence by \eqref{092}$_2$  
 $\mathbf{Z}_{0}^{\left( 1\right) }\mathbf{a}^{\left( 1\right) }=\mathbf{0,}
$ i.e. $\mathbf{a}^{\left( 1\right) }$ is the null vector of $\mathbf{Z}%
_{0}^{\left( 1\right) }$ for any anisotropy. 
This property, combined with \eqref{111}, implies that $\mathbf{Z}_{0}^{\left( 1\right) }$ has the structure
\beq{503}
\mathbf{Z}_{0}^{\left( 1\right) }
=  \begin{pmatrix}
a&ia&c\\
-ia&a&-ic\\
c^*&ic^*& b
\end{pmatrix}  
\text{  with   }
 \begin{pmatrix}
a ~  & c \\
 & \\
 c^* &~ b 
\end{pmatrix} 
 \text{  negative definite}. 
\eeq
The $2\times 2$ matrix becomes diagonal $(c=0)$ for  symmetry as low as monoclinic and   an explicit 
 form of $\mathbf{Z}_{0}^{\left( 1\right) }$ can then be   found, 
see \S\ref{7.1.1}.

\subsection{The matrix $ {\mathbf Z}_0\o$ at large azimuthal order $n$}

For $n\gg 1$ we assume an asymptotic expansion of the impedance in inverse powers of $n$: 
\beq{94}
 {\mathbf Z}_0\o = n {\mathbf z}_{0} + {\mathbf z}_{1} + n^{-1}{\mathbf z}_{2}+ \ldots, 
 \eeq
 where ${\mathbf z}_{0}$, ${\mathbf z}_{1}$, \ldots, are independent of $n$. 
Substituting into \eqref{25} and comparing terms of like powers in $n$ yields a sequence of matrix equations, the first  of which is
\beq{020}
\big( {\mathbf z}_{0} - i {\mathbf R}_0^T\big) {\widehat{\mathbf Q}_0}^{-1} 
\big( {\mathbf z}_{0} + i {\mathbf R}_0\big) -\widehat{\mathbf T}_0  = {\mathbf{0}} . 
\eeq
This  algebraic Riccati equation can be identified as eq.\  \eqref{--2} with system matrix ${\mathbf Q}= i k_\theta {\mathbf N}$
where  $k_\theta = n/r$ and ${\mathbf N}$ is the (static)  Stroh matrix for   the sagittal plane defined by ${\mathbf e}_r,\, {\mathbf e}_\theta = {\mathbf n},\, {\mathbf m}$. 
The subsequent identities are inhomogeneous Lyapunov equations
\beq{0488}
{\mathbf E}^+  {\mathbf z}_{j} + {\mathbf z}_{j}{\mathbf E} +{\mathbf f}_j({\mathbf z}_{0},{\mathbf z}_{1}, \ldots {\mathbf z}_{j-1}\big) ={\mathbf{0}}, \quad j=1,2,\ldots,
\eeq
with the constant matrix operator 
\beq{-55}{\mathbf E}
= {\widehat{\mathbf Q}_0}^{-1} 
\big( {\mathbf z}_{0} + i {\mathbf R}_0\big), 
\eeq
where  
\beq{-56}
{\mathbf f}_1 = i \widehat{\mathbf T}_0{\mathbf K} + i{\mathbf K}\widehat{\mathbf T}_0 +
{\mathbf E}^+  {\mathbf R}_{0}{\mathbf K} + {\mathbf K}{\mathbf R}_{0}^T {\mathbf E}, \quad \text{etc.}
\eeq

The leading order impedance  ${\mathbf z}_{0}$ is the solution of the matrix algebraic Riccati equation \eqref{020}, and may be determined by the  methods discussed above (via the eigenvectors \& eigenvalues, or the integral representation).   Subsequent terms ${\mathbf z}_{j}$, $j=1,2,\ldots$, 
satisfy a Lyapunov equation \eqref{0488} with different right hand sides but  the matrix Lyapunov operator  is the same in each case.   
The solution of the Lyapunov equation depends upon the spectrum of ${\mathbf E}$, and since the eigenvalues of 
${\mathbf E}$ have negative real part, it follows that the unique solution is 
\beq{235}
{\mathbf z}_{j}
=   
\int\limits_0 ^{\infty}  \dd s\, \ee^{s  {\mathbf E}^+} 
{\mathbf f}_j 
\ee^{s  {\mathbf E}} .
\eeq
The asymptotic sequence in inverse powers of $n$ can thus be evaluated to any desired order.

\section{Radiation impedance matrix }\label{sec6}

 The radiation impedance is relevant, for instance, in a configuration of infinite outer extent in which the cylinder is inhomogeneous in $r< r_0$ for finite  $r_0$ and uniform otherwise. 
It is always possible to split the linear total field into incident and scattered components, such that the scattered solution in $r>r_0$ has 
only  positive radial  energy flux.  The radiation impedance is   defined by  the subset of wave solutions with this radiation property.   
 
\subsection{Explicit form of  ${\mathbf Z}\o_\mathrm{rad}(r)  $}\label{sec6.1}

An alternative partitioning of the integral matrix is required to account for the separate  radiating, i.e. non-zero flux, modes.  This may be accomplished by a change of   basis   that brings about  the diagonal form of the flux matrix $\pmb{\mathcal{N}}^{+}\left( r\right) \mathbf{T}\pmb{\mathcal{N}}\left( r\right)$ \cite{Shuvalov03}.  
Proceeding from the integral matrix $ \pmb{\mathcal{N}}$ which satisfies 
\eqref{885}, and hence composed of modes with zero radial flux, we first 
convert ${\mathbf T} $ to diagonal form by an orthogonal transformation, 
\beq{757}
{\mathbf T}  = {\mathbf W} {\mathbf J}{\mathbf W}^+, 
\quad
{\mathbf J}  =  \begin{pmatrix} -{\mathbf I} & {\mathbf 0} \\ {\mathbf 0} & {\mathbf I} \end{pmatrix},
\quad
{\mathbf W} = \frac1{\sqrt{2}} 
\begin{pmatrix} {\mathbf I} & {\mathbf I} \\ -{\mathbf I} & {\mathbf I} \end{pmatrix} 
\quad ({\mathbf W}^+{\mathbf W} = {\mathbf I} ).
\eeq
Then referring to the notation \eqref{6}$_1$ for $ \pmb{\mathcal{N}}$ satisfying 
eq.\ \eqref{885},  we are led to  the following partitioning of the integral matrix,
\beq{-459}
  \pmb{\mathcal{N}}_1\left( r\right) =\pmb{\mathcal{N}}\left( r\right){\mathbf W} = 
  \begin{pmatrix}
\widehat{\mathbf{U}}_{+} & \widehat{\mathbf{U}}_{-} \\ & \\
\widehat{\mathbf{V}}_{+} & \widehat{\mathbf{V}}_{-}%
 \end{pmatrix}= 
 \frac1{\sqrt{2}} \begin{pmatrix}
\widehat{\mathbf{U}}_{1} -\widehat{\mathbf{U}}_{2} & \widehat{\mathbf{U}}_{1}  + \widehat{\mathbf{U}}_{2} \\ & \\
\widehat{\mathbf{V}}_{1} - \widehat{\mathbf{V}}_{2} & \widehat{\mathbf{V}}_{1}+\widehat{\mathbf{V}}_{2}%
 \end{pmatrix}. 
\eeq  
The $+$ and $-$ suffices indicate modes that have positive and negative flux in the radial direction, which is evident from the sign of the flux defined by eq. \eqref{010}, and 
the flux condition \eqref{885} which becomes  
\begin{equation}
 \pmb{\mathcal{N}}^{+}_1\left( r\right) \mathbf{T}\pmb{\mathcal{N}}_1\left( r\right)  
=  {\mathbf J}   \quad \text{for }n>1
\qquad 
\big( \Rightarrow \,  \pmb{\mathcal{N}}_1\left( r\right) \mathbf{J}\pmb{\mathcal{N}}_1^{+}\left( r\right)  
=  {\mathbf T} \big)
.
\label{8-7}
\end{equation}
Extension of this identity to the  special cases $n=0$, $1$, is contingent on the details of the Frobenius solutions, see \S\ref{4.1.2}. 
In the cases of  transversely isotropy and  isotropy, the 
  $+$ and $-$ modes correspond to radiating (outgoing) and incoming Hankel function solutions, 
$H^{(1)}_n$ and $H^{(2)}_n$ respectively.

The wave-based partition \eqref{-459} provides the required  modes to express the radiation impedance,  defined in \eqref{+6}, with $ \mathbf{U}_\mathrm{scat}, \mathbf{V }_\mathrm{scat}
\rightarrow \widehat{\mathbf{U}}_{+} , \widehat{\mathbf{V}}_{+} $, 
\beq{637}
{\mathbf Z}\o_\mathrm{rad}(r) = i 
\widehat{\mathbf{V}}_{+} 
{\widehat{\mathbf{U}}_{+}}^{-1} , \quad r> 0.
\eeq
It is important to note that the radiation impedance is not Hermitian, since according to \eqref{8-7}
\beq{99-}
{\mathbf Z}\o_\mathrm{rad} - {\mathbf Z}\o_\mathrm{rad}^+ = -i \, \big( 
{\widehat{\mathbf{U}}_{+}}{\widehat{\mathbf{U}}_{+}}^+ 
\big)^{-1} \ne 0, 
\eeq
 which implies in fact that  $i ({\mathbf Z}\o_\mathrm{rad} - {\mathbf Z}\o_\mathrm{rad}^+)$ is Hermitian and positive definite. 
 
 As an example, consider SH wave motion in a uniform  isotropic solid with $k_z=0$, for which the scalar radiation impedance is
$Z_\mathrm{rad}(r) =     -c_{44} \, k r   {H^{(1)'}_n (k r)}/{H^{(1)}_n(k r)}$ 
where $k=\omega \sqrt{\rho /c_{44} }$ and $H^{(1)}_n$ is the Hankel function of the first kind, 
see eq. \eqref{33}.  Using known properties of cylindrical functions yields for this case 
\beq{4--3}
i( Z_\mathrm{rad} - Z_\mathrm{rad}^+)
= 4\pi^{-1}c_{44} |H^{(1)}_n(k r)|^{-2} >0. 
\nonumber 
\eeq  
Note that for large values of $kr$ the SH radiation impedance is $Z_\mathrm{rad}(r) = -ikr c_{44} + \frac12 c_{44} +$O$\big( (kr)^{-1}\big)$.

\subsection{Asymptotic form of  ${\mathbf Z}\o_\mathrm{rad}(r)  $ as $r\rightarrow \infty $ }

Assume that the cylinder material is homogeneous for $r_0< 
r < \infty $, for some finite radius $r_0$.  As $r\rightarrow \infty$  the impedance ${\mathbf Z}\o_\mathrm{rad}(r)$, which we recall is defined with generalized traction vector ${\mathbf V} = ir\mathbf{\Upsilon}\o(r)$, may grow without bound while   $r^{-1}{\mathbf Z}\o_\mathrm{rad}(r)$ tends to a  planar limit.   This behaviour is evident for  the SH radiation impedance considered 
in \S\ref{sec6.1} which is proportional to $r$ as $r\rightarrow \infty$.   We therefore 
assume that the radiation impedance has the form 
\beq{638}
{\mathbf Z}\o_\mathrm{rad}(r) = k_z r \overline{\mathbf Z}\o_{\infty} +\text{O}(1), 
\quad r\rightarrow \infty , 
\eeq
where   $k_z \overline{\mathbf Z}\o_{\infty}$  is a constant matrix.  
 This can be found by 
considering the large $r$ limit of  the differential system \eqref{130}, which  reduces to its plane wave asymptote with $r$ playing the role of a rectangular coordinate: 
\beq{230}
  \frac{\dd}{\dd r}   \pmb{\phi}\o(r)=  i\,\mathbf{f}_0 \pmb{\phi}\o(r),
 \eeq
where $\pmb{\phi}\o(r)$ is a $6-$vector and $\mathbf{f}_0$ a  matrix constant
\cite{Shuvalov03} 
\beq{260}
 \pmb{\phi}\o(r)= 
\begin{pmatrix}
{\mathbf{U}\o(r)}
\\
{\overline{\mathbf{V}}\o(r)}
\end{pmatrix}, 
\quad 
{\overline{\mathbf{V}}\o(r)}=
i k_z^{-1}\mathbf{\Upsilon}\o(r),
\qquad 
 \mathbf f_0= k_z \begin{pmatrix} {\mathbf g_1^{\{1\}}}&{\mathbf g_0^{\{2\}}}
\\ & \\ 
k_z^{-2} {\mathbf g_2^{\{3\}}}& {\mathbf g_1^{\{1\}+}}
\end{pmatrix} .  
\eeq 
The six independent solutions to \eqref{260} may be separated into  triplets  according to their flux properties, with $\mathbf{U}_+, \overline{\mathbf{V}}_+$ signifying the outgoing, or radiating solutions. 
The limiting  radiation impedance is  then defined by analogy with \eqref{637} as 
\beq{-45=}
\overline{\mathbf Z}\o_{\infty} = i \overline{\mathbf{V}}_+   \mathbf{U}_+^{-1}. 
\eeq

Properties of $\overline{\mathbf Z}\o_{\infty}$ can be deduced by noting that the system \eqref{230} is equivalent to that for  a half-space with the identification  $\mathbf{f}_{0}= k_{z}\mathbf{N}\left( v\right) $, where $v = \omega /k_z$ and $\mathbf{N} \left( v\right) $ is the elastodynamic Stroh matrix for the sagittal plane 
$\{ {\mathbf e}_r,\, {\mathbf e}_z\} = \{{\mathbf n},\, {\mathbf m}\}$. 
This enables us to equate   the limiting radiation   matrix $ \overline{\mathbf Z}_\infty$   with the surface impedance matrix $\mathbf{Z}\left( v\right) $  for a homogeneous half-space \cite{Barnett85}.   Consequently, 
 $\overline{\mathbf Z}\o_{\infty }= \overline{\mathbf Z}\o_{\infty }^+$ for subsonic $v$, 
 i.e.  $0\le v \le \hat{v}$.  
  The possibility of 
 $\overline{\mathbf Z}\o_{\infty }$ being  Hermitian 
 seems at odds with the conclusion \eqref{99-}; however, it should be borne in  mind that  $\overline{\mathbf Z}\o_{\infty }$ is only the leading order term in the asymptotic series implicit in \eqref{638}.    The subsonic situation  may be understood in the context of the  SH radiation impedance example above with the wavenumber $k$ formally taken as  imaginary, in which case the  Hankel function  is replaced with the modified Bessel function of the second kind via the identity $H^{(1)}_n(x) = 2\pi^{-1}(-i)^{n+1}K_n(-ix)$.  
 Conversely, $\overline{\mathbf Z}\o_{\infty }$  is not Hermitian for  $v>\hat{v}$  \cite{Barnett85}.   
The equivalence with the half-space problem also implies that $\overline{\mathbf Z}\o_{\infty }$ is a solution of the  algebraic matrix Riccati equation 
\beq{125}
  \big(  \overline{\mathbf Z}\o_{\infty} -i   {\mathbf P}_c^T \big)
 {\mathbf{\widehat{Q}}}_c^{-1} \big(  \overline{\mathbf Z}\o_{\infty} + i   {\mathbf P}_c  \big) 
  -  \widehat{\mathbf M}_c + \rho_cv^2{\mathbf I} ={\mathbf{0}} ,
\eeq
where the suffix $c$ indicates the  constant values in $r>r_0$.  Equation \eqref{125} can be deduced by analogy  with  eq.\ \eqref{020},
noting  the presence of the additional dynamic term $\rho_cv^{2}\mathbf{I}$
in $\mathbf{f}_{0}$ and hence in \eqref{125}. 
The Riccati equation indicates that as $k_z\rightarrow 0$ the matrix  $k_z\overline{\mathbf Z}\o_{\infty }\rightarrow \overline{\mathbf Z}\o_{\infty 0}  $ where  
$   \overline{\mathbf Z}\o_{\infty 0}  
 {\mathbf{\widehat{Q}}}_c^{-1}   \overline{\mathbf Z}\o_{\infty 0}  
   = -   \rho_c\omega^2{\mathbf I} $,  with  a unique solution  satisfying \eqref{99-}, and hence 
 \beq{0789}
\lim_{r\rightarrow \infty} r^{-1}{\mathbf Z}\o_\mathrm{rad}(r)  = - i \omega \rho_c^{1/2} {\mathbf{\widehat{Q}}}_c^{1/2} \quad\text{for }k_z=0. 
 \nonumber
 \eeq
Note that taking ${\mathbf{\widehat{Q}}}_c$  with $c_{15}$, $c_{56}=0$ and $c_{44}=c_{55}$  factors out the asymptotic form $Z_\infty =-ikr c_{44}$ of the above-mentioned scalar radiation impedance $Z_\mathrm{rad}$ for the SH waves in an isotropic solid.

Finally, it is emphasized  that developments in this subsection are irrelevant to the
solid-cylinder impedance $\mathbf{Z}\left( r\right) $ which, by
construction, is Hermitian at any $r$ and for any $v (=\omega/k_z)$.  This in fact  implies that 
  $r^{-1}\mathbf{Z}\left( r\right) $ cannot become   constant   as $%
r\rightarrow \infty $ because otherwise the  arguments  subsequent  to eq.\ \eqref{638}  would violate the unconditional Hermiticity of $\mathbf{Z}\left( r\right) $. 
For instance, the   out-of-plane  impedance, 
 $Z\o_z(r,0) =     -c_{44} \, k_2 r   {J_n'(k_2 r)}/{J_n(k_2 r)} $, see  eq.\   \eqref{33}$_2$, 
 has no   large-$r$ limit. 

\section{Explicit examples of the solid   impedance  }\label{sec7}

The central impedance ${\mathbf Z}_{ 0}\o $ is first presented for several cases of material symmetry, including monoclinic and orthorhombic.   
A semi-explicit form for ${\mathbf Z}\o (r)$ is possible if the material is transversely isotropic, providing a check on the numerical  calculations in \S\ref{sec7.3}.

\subsection{The central impedance ${\mathbf Z}_{ 0}\o $ }

It follows from its definition through ${\mathbf g}_0(0)$   that 
${\mathbf Z}_{ 0}\o $  depends at most on  15 of the 21 possible elastic moduli.  The six redundant moduli are those with suffix $3$ occurring in the Voigt notation.

\subsubsection{Monoclinic  symmetry}\label{7.1.1}
For monoclinic symmetry with the  symmetry plane orthogonal to the $z-$axis the impedance has the structure
\beq{80}
{\mathbf Z}_0\o  =
\begin{pmatrix}
\quad {\mathbf Z}_{\perp 0}\o \quad &  \begin{matrix} 0 \\ 0 \end{matrix}
\\
0 \quad  0 & Z_{z0}\o
\end{pmatrix} ,
\eeq
where ${\mathbf Z}_{ \perp 0}\o$ and $Z_{z0}\o$ are  the in-plane and out-of-plane impedances, respectively.  The out-of-plane scalar impedance  follows from \citep[Eqs.\ (37), (38)]{Shuvalov03a} as
\beq{81}
Z_{z0}\o = - n \sqrt{c_{44}c_{55}- c_{45}^2}.
\eeq

For $n>1$ the in-plane  impedance can be expressed in semi-explicit form in terms of   the 
eigenvalues  $\lambda_j$, $\Re \lambda_j >0$, $j=1,2,$    of the $2\times 2$ matrix  ${\mathbf g}_{\perp 0}(0)$ formed from the upper left block of ${\mathbf g}_{ 0}(0)$.   By use of the following identity  for $2\times 2$ matrices,
\beq{99}
{\pmb \lambda}   + \lambda_1 \lambda_2{\pmb \lambda}^{-1}
= ( \lambda_1 +\lambda_2 ){\mathbf I} \quad (\lambda_1 \lambda_2 \ne 0),
\nonumber
\eeq 
the formulae in \eqref{74} may be combined to eliminate the explicit dependence on the eigenvector matrix, with the result  
\beq{749}
{\mathbf Z}_{\perp 0}\o
= - \frac12 (\widetilde{\mathbf R}_0 + \widetilde{\mathbf R}_0^+)
- \big(  {\mathbf{\widehat{Q}}}_0^{-1} + \lambda_1 \lambda_2 \widetilde{\mathbf T}_0^{-1}  \big)^{-1}
\big[ \big(  {\mathbf{\widehat{Q}}}_0^{-1} - \lambda_1 \lambda_2 \widetilde{\mathbf T}_0^{-1}  \big) \frac12 
 (\widetilde{\mathbf R}_0 - \widetilde{\mathbf R}_0^+)
+ (\lambda_1 + \lambda_2){\mathbf I}\big]
. \nonumber 
\eeq
Note that the matrices on the right hand side are all $2\times 2$, i.e. 
$\widetilde{\mathbf R}_0 = \widetilde{\mathbf R}_{\perp 0}$, etc., and the eigenvalues are the two roots of the quartic  
$\det {\mathbf \Lambda}_\perp  = 0$ from eq.\ \eqref{77} with positive real parts. 
The block impedance ${\mathbf Z}_{\perp 0}$  depends upon the six in-plane moduli. 
 $c_{\mu\delta}$ $(\mu,\, \delta  
 = 1,2,6)$.  

For $n=1$ the in-plane  impedance  possesses  a null vector as described in \S \ref{5.3.3}, and based on the required Hermiticity, it must   have  the form
\beq{-33}
{\mathbf Z}_{\perp 0}^{(1)}   
= z^{(1)} \begin{pmatrix}
1 & i \\
-i & 1 
\end{pmatrix}
= z^{(1)}\,  {\mathbf e}^+{\mathbf e}, 
\qquad
{\mathbf e} =  \big( 1 ,\,  i \big).
\eeq
The algebraic Riccati equation  \eqref{25} then reduces  to  
\beq{-25}
 \big\{ \big(  z^{(1)} {\mathbf e}  -i {\mathbf e}  {\mathbf R}_0^T \big)
 {\mathbf{\widehat{Q}}}_0^{-1} \big(   z^{(1)} {\mathbf e}^+ +  i {\mathbf R}_0 {\mathbf e}^+   \big) 
  - {\mathbf e}{\mathbf{\widehat T}}_0{\mathbf e}^+ \big\} {\mathbf e}^+{\mathbf e}  ={\mathbf{0}} ,
  \nonumber 
\eeq
implying a quadratic equation for $ z^{(1)}$,
\begin{align*}
& z^{\left( 1\right) ^{2}}\left( c_{11}+c_{66}\right) -2z^{\left( 1\right)
}\left( c_{11}c_{66}-c_{16}^{2}-c_{12}c_{66}+c_{16}c_{26}\right) 
\\ & \qquad 
-\left(
c_{11}c_{22}c_{66}+2c_{12}c_{16}c_{26}-c_{11}c_{26}^{2}-c_{22}c_{16}^{2}-c_{12}^{2}\allowbreak c_{66}\right) =0.%
\end{align*}
The unique physical $ z^{(1)}$ is, according to \S\ref{sec5.25},  provided by the negative  root. We note that 
the eigenvalues for the in-plane modes are $\lambda _{1}^{\left( 1\right)
}=0$   and $\lambda _{2}^{\left( 1\right) }$ that is the physical (positive real part)
root of 
\begin{equation}
\lambda ^{2}\left( c_{11}c_{66}-c_{16}^{2}\right) +2i\lambda \left(
c_{11}c_{26}-c_{12}c_{16}\right) +\left( c_{16}-c_{26}\right)
^{2}+c_{66}\left( 2c_{12}-c_{11}-c_{22}\right) -c_{11}c_{22}+c_{12}^{2}=0.
\label{vp}
\end{equation}

For  $n=0$ the general expression   \eqref{3-} reduces to  
\beq{5053}
z^{\left( 0\right) }= -c_{12}+\frac{c_{16}c_{26}}{c_{66}}-\sqrt{\left( c_{11}-%
\frac{c_{16}^{2}}{c_{66}}\right) \left( c_{22}-\frac{c_{26}^{2}}{c_{66}}%
\right) }. \nonumber 
\eeq

\subsubsection{Orthorhombic and tetragonal  symmetry}
For the orthorhombic symmetry and $n>1,$ the in-plane impedance $\mathbf{Z}_{\perp 0}^{\left( n\right) }$ is given by the upper 2$\times $2 block of
\eqref{74}:
\beq{803}
{\mathbf Z}_{\perp 0}\o
= \begin{pmatrix}
-c_{12}  & -in   c_{12} 
\\  
-in   c_{66} &  c_{66} 
\end{pmatrix}
-
\begin{pmatrix}
   c_{11}  & 0
\\ 
0 &     c_{66} 
\end{pmatrix}
{\mathbf A}_{\perp } 
\begin{pmatrix}
\lambda_1  & 0
\\  
0 & \lambda_2
\end{pmatrix}
{\mathbf A}_{\perp }^{-1},\quad n> 1,
\eeq
where  $\lambda_{1,2}$ are the physical roots of the equation
\beq{088}
\lambda ^{4}c_{11}c_{66}-\lambda ^{2}\left[ n^{2}\left(
c_{11}c_{22}-c_{12}^{2}-2c_{12}c_{66}\right) +c_{66}\left(
c_{11}+c_{22}\right) \right] +c_{22}c_{66}\left( n^{2}-1\right) ^{2}=0, 
\nonumber
\eeq
and  ${\mathbf A}_{\perp }
=\left\Vert \mathbf{a}_{1\perp },\mathbf{a}_{2\perp }\right\Vert $ is composed of  
the null vectors of the matrix 
$\mathbf{\Lambda }_{\perp }\left( \lambda \right) $, and  can be expressed
\beq{082}
{\mathbf A}_{\perp } =
\begin{pmatrix} 
\lambda_1^2c_{66}  - c_{66} - n^2 c_{22}  & -in [  \lambda_2 (c_{12}+c_{66}) -  c_{22} - c_{66}]  
\\  & \\  
-in [ \lambda_1 (c_{12}+c_{66}) + c_{22} + c_{66} ] & \lambda_2^2c_{11} - c_{22} - n^2 c_{66} 
\end{pmatrix}.\nonumber
\eeq
 
 For $n=1$ the scalar in-plane impedance is 
 \begin{equation*}
z^{\left( 1\right) }=\frac{c_{66}}{c_{11}+c_{66}}\left(
c_{11}-c_{12}-c_{11}\lambda _{2}^{\left( 1\right) }\right),
\end{equation*}%
where $\lambda _{2}^{\left( 1\right) }=\sqrt{ { 
(c_{11}c_{22}-c_{12}^{2}+c_{11}c_{66}+\allowbreak c_{66}c_{22}-2c_{12}c_{66} )}/{ (
c_{11}c_{66} )}}$ is the (physical) root of (\ref{vp}) simplified for the
orthorhombic case.

For tetragonal symmetry with $c_{16}=c_{26}=0$ the in-plane impedance is unchanged from \eqref{803},  and the out-of-plane impedance \eqref{81} further simplifies due to $c_{44}=c_{55}$ (on top of the orthorhombic condition $c_{45}=0)$.

\subsubsection{Transverse isotropy and isotropy}

The central-impedance matrix  reduces for transversely isotropic symmetry  to 
\bal{34}
{\mathbf Z}_0\o &=
\begin{pmatrix}
2 c_{66} \big( \frac{c_{66} - n c_{11} }{ c_{66} + c_{11}}  \big)
  &  i2 c_{66} \big(\frac{ n c_{66} -   c_{11} }{  c_{66} + c_{11}} \big)  &0
\\ & \\
-i2 c_{66} \big(\frac{ n c_{66} -   c_{11} }{  c_{66} + c_{11}} \big)  &
2 c_{66} \big( \frac{c_{66} - n c_{11} }{ c_{66} + c_{11}}  ) & 0
\\ && \\
0& 0& -n c_{44}
\end{pmatrix} , \quad n\ne 0,
\nonumber \\ & \\
 {\mathbf Z}_0^{(0)} &= -2( c_{11} - c_{66}) 
\begin{pmatrix}
1  &  ~0~  &0
\\ 
0  &0 & 0
\\ 
0& 0& 0
\end{pmatrix} ,
\nonumber 
\end{align}
which applies, of course, to isotropy  $(c_{44} = c_{66})$.   Equation \eqref{34} is also derived in the next subsection. 

\subsection{The solid-cylinder   impedance  ${\mathbf Z}(r)  $ for transverse isotropy}

\subsubsection{General formulation}\label{7.2.1}

The constitutive relation for ${\mathbf t}_r$   
combined with  eqs.\ \eqref{10} and \eqref{901}$_3$, implies   for any material anisotropy, 
\beq{001}
{\mathbf Z}\o(r) = -\widetilde{\mathbf R}-ik_z r{\mathbf P}  - {\mathbf{\widehat{Q}}}
 \big( r\frac{\dd}{\dd r} \widehat{\mathbf U}_1\big) \,  \widehat{\mathbf U}_1^{-1} 
,\quad r\ge 0,  
\eeq 
 where the matrix $\widehat{\mathbf U}_1 (r)$ is  any unnormalized triad of independent physical solutions.  The radiation impedance is obtained if the matrix is replaced with $\widehat{\mathbf U}_+ (r)$ comprising  linearly independent radiating solutions. 
The difficulty in applying \eqref{001} is that explicit matrix solutions for $\widehat{\mathbf U}_1$ or $\widehat{\mathbf U}_+$  are not generally available  except under certain restrictions on material symmetry, such as transverse isotropy. 

Assuming transverse isotropy, solutions for the displacements that are either regular at $r=0$ or radiating to infinity can be constructed in terms of cylinder function by adopting the  representation of \citet{Buchwald59} (see also  \cite{Rahman98}). Thus,  
\beq{333}
\widehat{\mathbf U}\o(r) = \begin{pmatrix}
C_n'(k_1r)    &   C_n'(k_2r)  & -\frac{in}{k_3r} C_n(k_3r)
\\  & & \\
\frac{in}{k_1r} C_n(k_1r)     & \frac{in}{k_2r} C_n(k_2r) &  C_n'(k_3r)
\\  & & \\
i\frac{\kappa_1}{k_1} C_n(k_1r)  & i\frac{\kappa_2}{k_2} C_n(k_2 r) & 0 
\end{pmatrix} , \nonumber
\eeq 
where the principal wavenumbers $k_1$,  $k_2$, $k_3$, and auxiliary wavenumbers 
$\kappa_1$,  $\kappa_2$, are 
\begin{align}
k_{1,2}^2 &=
\frac{a\mp \sqrt{a^2- b}}{2c_{11}c_{44}},
\qquad
k_3^2 = \frac{\rho \omega^2 - c_{44} k_z^2}{c_{66}},
\quad   \kappa_i =  \frac{c_{66}k_3^2-c_{11}k_i^2  } { k_z(c_{13}+c_{44})} \, \, \,\, (i=1,2) ,
\nonumber \\ 
 a &=( c_{11}+c_{44})\rho \omega^2+
 ( c_{13}^2 +2c_{13}c_{44} -c_{11}c_{33} )
 k_z^2  , \quad
 b =4c_{11}c_{44} (\rho \omega^2-c_{33}k_z^2)(\rho \omega^2-c_{44}k_z^2) ,
 \nonumber
\end{align}
and  $C_n = J_n$ for displacements regular at $r=0$ ,  $C_n = H_n^{(1)}$ for radiating solutions, 
 where $J_n$ are Bessel functions and $ H_n^{(1)}$ are Hankel functions of the first kind.  

Evaluating  \eqref{001} and simplifying terms using the identities $c_{44}\kappa_1\kappa_2 + c_{66}  k_3^2 =0$,  
$c_{11}( \kappa_1 k_2^2  - \kappa_2 k_1^2) =c_{66}k_3^2( \kappa_1   - \kappa_2  ) $,
we find that the solid cylinder impedance for transverse isotropy is 
\bal{+34}
 {\mathbf Z}\o(r) &=
\begin{pmatrix}
2c_{66}  & in 2c_{66} & ik_zr c_{44}
\\ 
-in 2c_{66} & 2c_{66} & 0
\\ 
-ik_zr c_{44} & 0 & Z\o_{z } 
\end{pmatrix}
+c_0 \begin{pmatrix}
\xi_3 (y_1-y_2)   & in (y_1-y_2)  & i\xi_3(\xi_1-\xi_2) 
\\ 
-in (y_1-y_2) & \xi_2 y_1 - \xi_1 y_2   & n(\xi_1-\xi_2) 
\\ 
-i \xi_3(\xi_1-\xi_2)  & n(\xi_1-\xi_2)  & 0
\end{pmatrix},
\\
 Z\o_{z } &= c_{44}\bigg(
\frac{n^2(\xi_1y_1-\xi_2y_2) - \xi_1\xi_2\xi_3 (y_1-y_2) }
     { \xi_3 (\xi_2y_1-\xi_1y_2) -n^2(y_1-y_2)  }
\bigg),
\quad
c_0 = \frac{c_{66} k_3^2r^2 } {  \xi_3 (\xi_2y_1-\xi_1y_2) -n^2(y_1  - y_2)},
\nonumber
\end{align}
with  the non-dimensional quantities 
\beq{335} 
y_i=\kappa_i r  \quad (i=1,2), \qquad 
 \xi_j = k_j r \frac{C_n'(k_j r)}{C_n(k_jr)} \quad (j=1,2,3) . \nonumber
\eeq 

The central-impedance limit may be extracted from \eqref{+34} by writing it in 
block form
\bal{+341}
 {\mathbf Z}\o(r,k_z) &=
 \begin{pmatrix}
\quad {\mathbf Z}\o_{\perp }(r,k_z)  \quad &  
		\begin{matrix} i(k_zr c_{44} +  c \xi_3 ) \\ nc \end{matrix}
\\
-i(k_zr c_{44} +  c \xi_3 ) \quad nc & Z\o_{z } (r,k_z)
\end{pmatrix} , \quad c = c_0 (\xi_1-\xi_2)  ,
\nonumber 
\\ &
\\ 
{\mathbf Z}\o_{\perp }(r,k_z) &=
2c_{66} \begin{pmatrix}
1 & in  
\\ & \\
-in  & 1
\end{pmatrix}
 + 
c_{66} k_3^2r^2
 \begin{pmatrix}
\frac{\xi_2 y_1 - \xi_1 y_2}{ y_1 - y_2}
& -in  
\\ & \\
in &
  \xi_3
\end{pmatrix}^{-1},
\nonumber
\end{align}
where the dependence 
on both $r$ and $k_z$ is   emphasized. 
For $k_z = 0$  we have
$k_j=\omega /c_j$ with $\rho c_1^2= c_{11}$, $\rho c_2^2= c_{44}$,  $\rho c_3^2= c_{66}$,
and  the impedance reduces to 
\beq{33}
\begin{split}
{\mathbf Z}\o(r,0)  & =
\begin{pmatrix}
\quad {\mathbf Z}\o_{\perp }(r,0)  \quad &  \begin{matrix} 0 \\ 0 \end{matrix}
\\
0 \quad  0 & Z\o_{z } (r,0)
\end{pmatrix} ,\quad\text{with     }
Z\o_z(r,0) =     -c_{44} \, k_2 r  \frac{C_n'(k_2 r)}{C_n(k_2 r)} ,
\\
{\mathbf Z}\o_\perp(r,0)  &=  2c_{66} \begin{pmatrix}
1 & in  
\\ & \\
-in  & 1
\end{pmatrix}
 + 
c_{66} (k_3 r)^2
 \begin{pmatrix}
 k_1 r \frac{C_n'(k_1 r) }{C_n(k_1 r)}
& -in  
\\ & \\
in &
  k_3 r \frac{C_n'(k_3 r) }{C_n(k_3 r)}
\end{pmatrix}^{-1}.
\end{split}
\eeq
 Taking the limit  $ r\rightarrow 0$ of \eqref{33} with the interior cylinder functions $C_n=J_n$  gives eq.\ \eqref{34}.

\subsection{Numerical example}\label{sec7.3}

\begin{figure}[t]
				\begin{center}	
				\includegraphics[width=4.3in , height=3.2in 					]{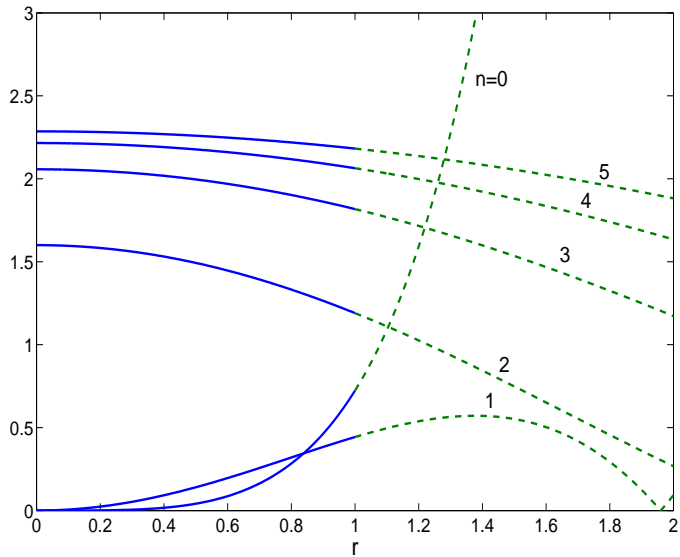} 
	\caption{The curves show $(n^3+1)^{-1} | \det {\mathbf Z}\o(r) |$ for $n=0,1,\ldots , 5$.  The material is isotropic with $\{ c_{11}, c_{66} , \rho\} = \{4,1,1\}$ and $\{\omega, k_z \} = \{ 1,0.2\}$. 
The Riccati equation \eqref{228} for ${\mathbf Z}\o(r)$ was integrated to obtain the curves for $0< r \le 1$, starting from $r=0$ with the  known ${\mathbf Z}\o_0$ of eq.\ \eqref{34}.  For $r>1$ the system equations \eqref{132} were integrated and eqs.\ \eqref{496} and \eqref{296} used to find ${\mathbf Z}\o(r)$, starting from the Riccati solution at $r=1$. }
		\label{fig2} \end{center}  
	\end{figure}

A procedure was outlined in \S\ref{4.3} for calculating the solid-cylinder impedance using  two separate  numerical solutions.  The Riccati equation \eqref{228} is first  integrated starting from $r=0$ with the  central impedance matrix ${\mathbf Z}_0$ as initial condition.   The integration proceeds up to $r=r_1$ where $r_1$ lies below   the first singularity of ${\mathbf Z}(r)$.  
For $r>r_1$ the impedance is obtained from \eqref{296}$_2$ as the solution of the the matricant-based system \eqref{496}, with  the Riccati solution at $r_1$ serving as the initial condition.  
As an illustration of its practicality, 
the two-stage algorithm was implemented  with representative  results  plotted in 
Figure \ref{fig2}.  

The initial step in the computation requires  the value of the 
 central impedance, which  was calculated   using  eqs.\ \eqref{-47} and \eqref{3-} for $n=0$  and  the  formula ${\mathbf Z}_0   = i  {\mathbf L}_1{\mathbf A}_1^{-1}$  for $n>0$, see \eqref{092}$_2$,  with  $ {\mathbf A}_1$, $ {\mathbf L}_1$ defined by the numerical spectral decomposition of ${\mathbf g}_0(0)$ and  the appropriate selection of its three   eigenvalues with positive real part.   
It was  confirmed that the computed ${\mathbf Z}_0$ satisfied the algebraic Riccati equation \eqref{25}, with error always less than   $10^{-12}$.  Numerical integration of equations \eqref{228} and  \eqref{496} was accomplished using the Runge-Kutta (4,5) routine in  Matlab. 
In order to assess the accuracy of the numerical results   the   computed matrix ${\mathbf Z}\o_{comp} (r)$ and the analytical solution for ${\mathbf Z}\o(r)$ of \eqref{+34} were compared. 
For the examples shown in Figure \ref{fig2} it was found  
that the     spectral norm 
of the difference satisfied  $ \|  {\mathbf Z}\o_{comp} (r) -{\mathbf Z}\o(r) \|_2 < 10^{-4}$ at all points.   The curves in  
Figure \ref{fig2}  use $r=1$ as the `cross-over' coordinate, but similar accuracy was found for other  values as long as they lie below  the first singularity of ${\mathbf Z}(r)$, which for the parameters considered is $r_*>2$.   In all cases the transition from the Riccati  to the matricant based solution was found to be smooth.   

This numerical procedure is designed to handle the coordinate-based singularity present in the system equations \eqref{130} at $r=0$, and can be continued, in principle,  to any finite $r$.  At the same time the computed impedance ${\mathbf Z}\o(r)$ will grow without bound at discrete values of $r>0$ associated with waveguide modes of the traction-free cylinder.  
  The point of the algorithm is that it will continue to provide accurate solution  regardless of the presence of two distinct types of singularity  at $r=0$ and at finite values.

 \section{Conclusion}\label{sec8}  
 
Impedance matrices appropriate to  cylindrically anisotropic radially inhomogeneous elastic materials
have been defined and procedures for their determination   developed.  In the process a new impedance matrix has been revealed as of central  importance for  wave motion in cylinders with  on-axis material.  The solid-cylinder impedance matrix is a characteristic property  of the cylinder, with no free parameters apart from     frequency and axial wavenumber.  The   impedance may be defined as the unique continuation of its  on-axis limit, the central-impedance matrix, which is  a   simpler  object  dependent only on (a subset of) the  elastic moduli.    Two methods have been described for   constructing the solid cylinder impedance at $r>0$, one  based on a Frobenius series solution, the other  using a differential Riccati equation.  In addition to providing practical means for computation, as has been demonstrated for the latter approach, the  methods  shed light on the structural properties of the impedances.  The Frobenius solution offers  direct proof of uniqueness and Hermiticity, while the Riccati solution provides a stable method to integrate the otherwise singular system of equations at $r=0$. 
  The radiation impedance matrix, suitable for infinite radial domains, has been defined and its properties   delineated.  We have found it  instructive to compare the cylindrical impedance matrices with the surface wave impedance for a homogeneous half-space.   The central-impedance matrix is the negative semi-definite counterpart of the static surface impedance, and  the large $r$ limit of the radiation impedance is  closely related to the surface wave impedance with $v=\omega/k_z$.

One  purpose in developing these impedance matrices is the significant advantage offered by the impedance approach in solving boundary value problems.   The solid-cylinder impedance matrix provides perhaps the simplest method to arrive at  the dispersion equation of  a radially inhomogeneous solid cylinder.  In this regard we note 
that, by analogy with the conditional $(3\times 3)$ and two-point ($6 
\times 6$) impedances of an annulus \cite{Shuvalov03}, the eigenvalues of the solid-cylinder
impedance should be monotonic in $\omega $ at any fixed $k_{z}$, which can be helpful for  finding   numerical solutions of  the dispersion equation.    
In a wider context, the impedance matrix  in conjunction with the radiation impedance, 
can  serve in formulating  scattering of acoustic and elastic waves from solid cylinders.  Other applications that we envisage include the use of impedance matrices for solving problems with distributed forces within the cylinder, and 
applications involving 2D-inhomogeneous or laterally bounded
planar and cylindrical waveguides \cite{Pagneux06,Getman90} where the  algebraic
impedance matrices discussed here  become differential  operators.

Another no less important reason for investigating the impedance matrix in the cylindrical context is that it affords new insights on the nature of   elastodynamic solutions in  anisotropic elasticity.   It is remarkable, for instance, to find the Riccati equation appear as a natural method for solution in cylindrical elastodynamics.  The Riccati equation, in fact, implies  that  the central-impedance  solves an algebraic Riccati equation, which in turn leads to direct methods for its evaluation  using   analogies with the surface wave impedance.   Differential  Riccati equations have been found useful in  few elastic wave settings, e.g. \cite{Biryukov85,Biryukov95,Caviglia02,Destrade09,Pagneux06}.  Its appearance here   suggests it has wider potential application in computational elastodynamics.

\bigskip 

\noindent \textbf{Acknowledgment.}  ANN wishes to express his gratitude to
the Laboratoire de M\'{e}chanique Physique (LMP) of the Universit\'{e}
Bordeaux 1 for their hospitality.
%

\end{document}